\documentclass[twocolumn,amsmath,amssymb,nofootinbib,superscriptaddress,longbibliography, floatfix]{revtex4-1}

\usepackage{amsmath,amssymb,amsfonts}
\usepackage{algorithmic}
\usepackage{graphicx}
\usepackage{textcomp}
\usepackage{xcolor}

\newcommand{\blue}[1]{{\color{blue} #1}}

\let\blue\relax

\newcommand{\qipacket}[1]{$\mathtt{#1}$}
\newcommand{\circlabel}[2][\relax]{${#1{\mathtt{#2}}}$}

\usepackage{hyperref}
\hypersetup{pdfstartview={FitH},pdfpagemode={UseNone}, breaklinks=true, colorlinks=true, linkcolor=blue, citecolor=blue, urlcolor=blue, bookmarksopen=true, pdfnewwindow=true}
\usepackage[all]{hypcap}
\usepackage[english]{babel}

\begin{document}

\title{A rechargeable AA battery supporting Qi wireless charging}

\author{Alexey~A.~Dmitriev}
\affiliation{School of Physics and Engineering, ITMO University, 49 Kronverksky pr., bldg. A, 197101 Saint Petersburg, Russia}

\author{Egor~D.~Demeshko}
\affiliation{School of Physics and Engineering, ITMO University, 49 Kronverksky pr., bldg. A, 197101 Saint Petersburg, Russia}

\author{Danil~A.~Chernomorov}
\affiliation{School of Physics and Engineering, ITMO University, 49 Kronverksky pr., bldg. A, 197101 Saint Petersburg, Russia}

\author{Andrei~A.~Mineev}
\affiliation{School of Physics and Engineering, ITMO University, 49 Kronverksky pr., bldg. A, 197101 Saint Petersburg, Russia}

\author{Oleg~I.~Burmistrov}
\affiliation{School of Physics and Engineering, ITMO University, 49 Kronverksky pr., bldg. A, 197101 Saint Petersburg, Russia}

\author{Sergey~S.~Ermakov}
\affiliation{School of Physics and Engineering, ITMO University, 49 Kronverksky pr., bldg. A, 197101 Saint Petersburg, Russia}

\author{Alina~D.~Rozenblit}
\affiliation{School of Physics and Engineering, ITMO University, 49 Kronverksky pr., bldg. A, 197101 Saint Petersburg, Russia}

\author{Pavel~S.~Seregin}
\affiliation{School of Physics and Engineering, ITMO University, 49 Kronverksky pr., bldg. A, 197101 Saint Petersburg, Russia}

\author{Nikita~A.~Olekhno}
   \email{nikita.olekhno@metalab.ifmo.ru}
\affiliation{School of Physics and Engineering, ITMO University, 49 Kronverksky pr., bldg. A, 197101 Saint Petersburg, Russia}

\date{\today}

\begin{abstract}
Wireless power transfer is one of the key drivers in modern consumer electronics, as it allows one to enhance the convenience and usability of many devices. However, in most cases, wireless charging is accessible only to devices with incorporated receivers or at least to gadgets with standard charging connectors, such as USB Type-C, that allow to attach an external receiver. We propose a rechargeable battery that has the size and output voltage of a standard AA battery but supports wireless power transfer from charging stations of the widely used Qi standard. The proposed design uses a series resonant circuit with a curved receiving coil, as well as load modulation using detuning capacitors switched by a microcontroller unit to implement a receiver compatible with the Qi Baseline protocol. It also utilizes a number of DC-DC converters to store energy in a Li-ion cell and convert it to the 1.5~V voltage level. Our design is supported by numerical simulations of magnetic field distributions and scattering parameters of the introduced battery coupled to a planar transmitting coil. The performance of the proposed battery has been studied experimentally, including measurements of the maximal distance between the battery and a charging station that allows wireless charging at various rotation angles and the charge curve. The developed battery design facilitates the addition of wireless charging functionality to a wide range of electronic devices in a universal way.
\end{abstract}

\maketitle

%_______________________Introduction__________________________
\section{Introduction}
\label{sec:Introduction}

Wireless power transfer (WPT)~\cite{2022_Yao_Simultaneous} finds numerous applications in areas where it solves technical issues related to the presence of wires, brushes, or connectors that provide power transmission, for example, in medical implants~\cite{2019_Mashhadi_New,2025_Wang_Design,2022_Roy_Powering,2025_Lee_Reconfigurable}, as well as in consumer devices where it allows one to enhance user comfort and device usability~\cite{2024_Alabsi}.

The key standard for magnetic resonant WPT for low- and medium-power consumer electronics is Qi~\cite{QiStandard}, which is widely used in smartphones, smartwatches, earphones, power banks, and computer mice, to name a few~\cite{2013_Hui,2015_Treffers_History,2020_Sousa_Optimal}. Qi transmitters are often integrated into public infrastructure and transport, with Internet of Things concepts being developed to optimize the usage of chargers~\cite{2019_Lai}. Moreover, many novel applications of Qi are actively being developed, ranging from health tracking rings~\cite{2021_Nguyen} to exotic examples such as electronic tattoos~\cite{2024_Kim_Unobstructive}, lab-on-a-chip centrifugal microfluidic platforms~\cite{2018_Delgado}, and swarms of tiny moving robots~\cite{2021_Li}. However, the majority of Qi receivers are currently incorporated into the devices being charged or connected to them via USB, limiting the applicability of wireless charging in peripherals not supplied with USB connectors.

At present, a substantial portion of portable electronic consumer devices, handheld tools, and toys rely on AA batteries, which, given their established infrastructure and utility, are expected to remain widespread for the next several decades~\cite{2023_Baicu_Embedded}. This has led to the development and adoption of Li-ion AA batteries rechargeable via USB and incorporating an 1.5~V DC-DC converter~\cite{2023_Frith}. However, the user experience is negatively impacted by the need to extract the rechargeable AA batteries from the consumer device for recharging.

%______________________Figure_1_______________________
\begin{figure}[b]
    \centering
    \includegraphics[width=7.5cm]{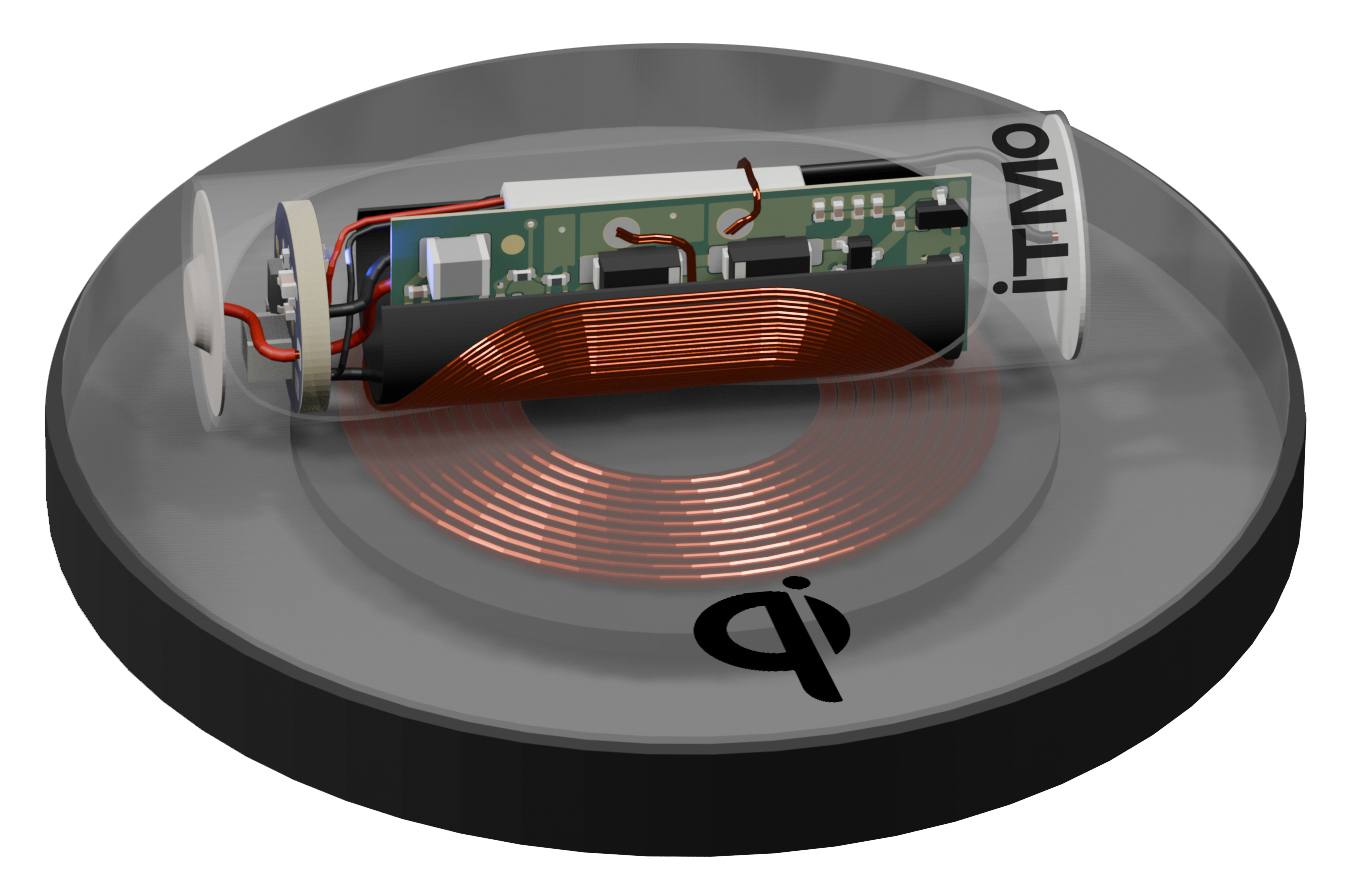}
    \caption{Design of the proposed battery. The AA battery-sized module which includes a receiving coil (copper), a receiver circuit board (green), a converter circuit board (blue), a rechargeable battery (gray), and a plastic casing (white) is placed atop the transmitting coil of a Qi standard charge station.}
    \label{fig:Concept}
\end{figure}
%______________________Figure_1_______________________

Therefore, the development of the AA battery that supports Qi wireless charging renders timely. A similar concept of a battery incorporating a Qi receiver has been implemented for the Apple Magic Mouse with a proprietary battery format~\cite{2016_Bidul}. A wirelessly rechargeable AA battery has also been demonstrated using electrodynamic WPT~\cite{2021_Smith}, however, its applicability is limited due to the necessity of a custom charging system.

As we demonstrate, the 14~mm diameter of AA batteries (which is small compared to standard Qi transmitting coils) limits the size of the receiving coil, resulting in weak coupling. Most Qi receivers use planar coils, but an estimate based on the Biot--Savart law immediately demonstrates that a 14~mm wide planar coil cannot achieve the necessary coupling with Qi transmitting coils for a power transfer contract to establish~\cite{Supplementary}. Therefore, the usage of a bent receiving coil appears necessary. Such bent coils have been studied as receivers~\cite{2020_Wen_Curvature} and transmitters~\cite{2023_Zhuang_Omnidirectional} for inductive and magnetic resonant WPT systems~\cite{2019_Sondhi,2022_Jeong}, but have not yet been applied in devices compatible with the Qi standard.

In the present paper, we introduce a rechargeable battery having the size and output voltage corresponding to those of an AA battery and incorporating a Qi wireless power receiver and a Li-ion cell, Fig.~\ref{fig:Concept}. The paper is organized as follows. In Sec.~\ref{sec:Structure}, we describe the structure of the proposed battery. In Sec.~\ref{sec:Implementation}, we describe the circuit implementation of each functional block, as well as the firmware executed by the microcontroller unit and responsible for forming the Qi packets. Then, in Sec.~\ref{sec:Simulations}, we discuss numerical simulations of a AA battery-sized curved receiving coil studying the dependence of the coupling strength on the curvature of the receiving coil, the distance between the receiving and transmitting coils, and the rotation angle of the receiving coil along the battery axis. Section~\ref{sec:Experiments} considers the realization and experimental tests of the battery prototype, including studies of the power transfer contract establishment and termination conditions, as well as the measurements of the charge curve. Section~\ref{sec:Outlook} contains a discussion of the results obtained and an outlook.

%______________________Figure_2_______________________
\begin{figure}[tb]
    \includegraphics[width=8.5cm]{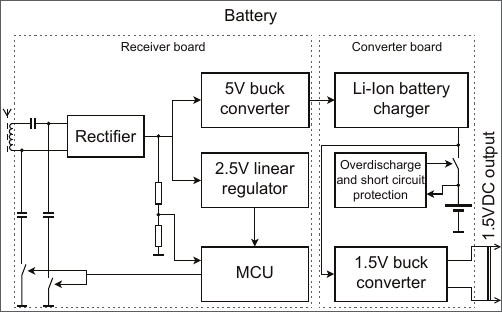}
    \caption{Structure of the proposed battery.}
    \label{fig:StructuralDiagram}
\end{figure}
%______________________Figure_2_______________________

%_________________________Schematics__________________________
\section{Structure of the battery}
\label{sec:Structure}

The proposed wirelessly charged battery includes the following components, depicted in Fig.~\ref{fig:Concept}: (i) a plastic casing having the shape of a cylinder with diameter $d=14.5$~mm and length $l=50$~mm corresponding to the dimensions of a standard AA battery; (ii) two metallic contact pads, the positive and negative terminals of the battery, at the top and bottom bases of the cylinder; (iii) Qi receiving coil which captures magnetic field created by a transmitting coil of a charging station; (iv) a receiver circuit board which rectifies the AC voltage from the receiving coil and converts it to $5$~V DC voltage; (v) a rechargeable Li-ion cell; and (vi) a converter circuit board responsible for proper charging of the Li-ion cell and converting its power to 1.5~V DC output voltage which is supplied to the positive and negative terminals of the battery.

The working principle of the proposed battery is illustrated by the structural diagram shown in Fig.~\ref{fig:StructuralDiagram}. First, the alternating magnetic field from the charging station, which has a frequency from 100 to 200~kHz, is converted into AC voltage by the receiving coil, which is a part of the input series resonant circuit tuned to the 100~kHz frequency. The output of this resonant circuit is connected to the rectifier, which can also be bypassed via two $22$~nF capacitors and switches driven by the microcontroller unit (MCU). Closing these switches creates another route for the AC current, rapidly changing the power drawn from the charging station, providing a means of load modulation, that is used in the Qi standard to establish the connection to the transmitter and provide digital feedback from the battery to the charging station via the Qi protocol~\cite{QiStandard,2010_VanWageningen}. The topology of the input resonant circuit with load modulation capacitors is chosen in accordance with the Qi standard.

%______________________Figure_3_______________________
\begin{figure*}[tbp]
   \centering
   \includegraphics{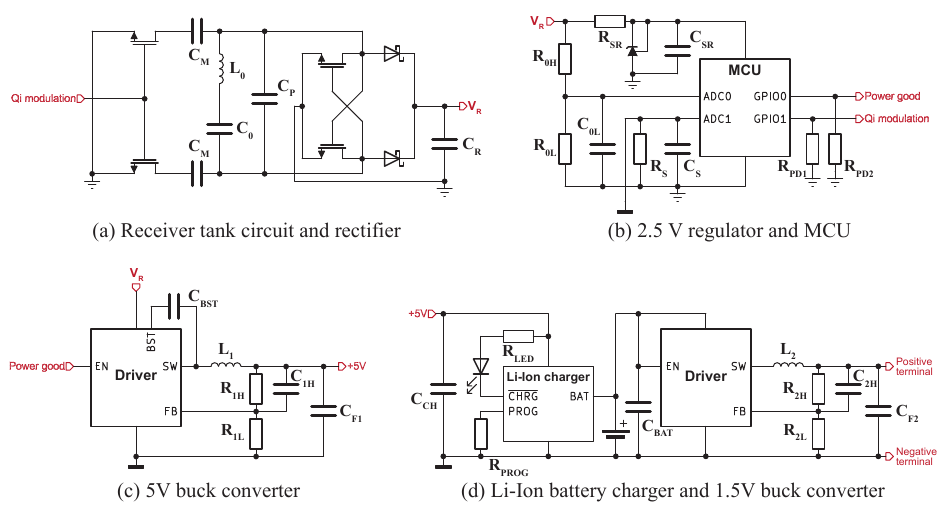}
   \caption{Circuit implementation of the proposed battery.}
   \label{fig:CircuitDiagram}
\end{figure*}
%______________________Figure_3_______________________

The DC voltage from the rectifier is distributed to the 5~V buck converter and to a low-power 2.5~V linear regulator providing the voltage supply to the MCU. MCU monitors the voltage at the rectifier output through a resistive divider and utilizes load modulation to send Qi packets to the charge station, which request it to transmit at a frequency closer or farther from the resonance, keeping the received voltage in the desired range (6 to 8 V). However, before a power transfer contract is established, the voltage at the rectifier output can reach 12~V, making the buck converter necessary to prevent overvoltage. MCU also monitors the current consumed by the battery, calculates the received power, and sends that information to the charging station, which constantly compares it with the transmitted power, providing foreign object detection as defined in the Baseline Qi protocol by terminating the power transfer if the difference between the transmitted and the received power exceeds a preprogrammed threshold.

The output of the 5~V buck converter is connected to a charging circuit, which performs the charging of the Li-ion cell in the constant-current (CC) or constant-voltage (CV) mode, depending on the state of charge of the cell. The cell itself is connected to the rest of the circuit through a protection unit based on a DW01A chip and 8205A double n-MOSFET, which monitors the cell voltage as well as the voltage across the open 8205A MOSFETs using their on-state resistance for current sensing, and disconnects the cell by closing the MOSFETs in case the cell gets overcharged, overdischarged, or short-circuited. The Li-ion cell is connected to the 1.5~V buck converter which converts its voltage to the 1.5~V level that is typical for AA batteries.

%__________________________Circuit_____________________________
\section{Battery circuit implementation}
\label{sec:Implementation}

The electric circuit of the proposed battery is implemented as a number of interconnected subunits, which are visible in Fig.~\ref{fig:Concept}: (i) Qi receiving coil; (ii) receiver circuit board comprising the input resonant circuit, the rectifier, the 5~V buck converter and the MCU as well as the 2.5~V linear regulator; (iii) a GoPower LP451124 65~mAh lithium polymer cell comprising a conventional protection unit based on the DW01A battery protection IC and a 8205A dual n-MOSFET; (iv) the converter circuit board comprising the Li-ion battery charger and the 1.5~V buck converter, and (v) positive and negative terminals of the battery. The receiver circuit is assembled on a rectangular printed circuit board made of FR4 with a 0.8~mm thickness and two-sided 18~\textmu m copper metallization; all components are mounted on the front side of the PCB. The converted circuit is assembled on a disk-shaped printed circuit board made of FR4 with a 1.6~mm thickness and two-sided 18~\textmu m copper metallization; the components are mounted on both sides of the PCB.

The receiving coil is a commercially available Qi-compatible planar coil in the shape of a rounded rectangle with length $40.5$~mm and width $30.5$~mm, consisting of $14$ turns of a wire made of two parallel monolithic enameled copper conductors with a circular cross section and a diameter of $0.4$~mm, which are soldered together at their ends, with an inner $22.5 \times 10.5~\text{mm}^2$ empty space. The coil is bent around a cylindrical surface with the axis parallel to the large side of the coil and the curvature radius $\rho=7$~mm. The receiving coil is backed with a $41 \times 28~\text{mm}^2$ flexible ferrite sheet with a thickness of $0.1$~mm from the interior side to provide a magnetic shielding that decouples other components of the battery from the transmitter. The resulting inductance $L_0$ of the coil is 12~\textmu H.

The implementation of the receiver resonant circuit, the capacitive load modulation subunit, and the rectifier is shown in Fig.~\ref{fig:CircuitDiagram}~(a). The capacitance $C_0$ of the resonant circuit is made of four 68~nF ceramic capacitors with a 25~V voltage rating connected in parallel to provide the 100~kHz resonant frequency of the input $LC$ circuit. The capacitor $C_{\text{P}}$ is required by the Qi specification for resonant receiver position detection at 1~MHz frequency~\cite{QiStandard}. The load modulation capacitors $C_{\text{M}}$ are connected to the reference ground by Si2302 n-MOSFET switches, which were chosen due to their low threshold voltage and on-state resistance. Both switches are closed simultaneously, as their gates are connected to the same wire (labeled as ``Qi modulation'' in Fig.~\ref{fig:CircuitDiagram}). The value of the load modulation $C_{\text{M}}$ was chosen in accordance with Power Receiver Design Example 3 from the Qi protocol specification~\cite{QiStandard}. All of the capacitors in the receiver circuit have a 25~V voltage rating, which is double the peak anticipated voltage.

The rectifier is implemented as a full bridge circuit, with two SK14 Schottky diodes comprising the positive half of the bridge and the intrinsic body diodes of Si2306 n-MOSFETs acting as the negative half of the bridge. The gates and drains of the MOSFETs are connected crosswise so that on the positive or negative half-cycle of the AC voltage, the gate voltage of the MOSFET, whose body diode is biased forward, becomes positive, opening that MOSFET and decreasing the rectifier voltage drop. This rectifier topology offers a lower voltage drop compared to a full bridge rectifier consisting of four Schottky diodes, see the Supplementary Material for comparison~\cite{Supplementary}. The rectifier uses a smoothing capacitance $C_{\text{R}}$. For the values of the components here and later, see Table~\ref{tab:ComponentValues}.

Fig.~\ref{fig:CircuitDiagram}~(b) shows the circuit implementation of the 2.5~V linear regulator and the MCU, which is responsible for establishing the receiver-to-transmitter feedback loop. The linear regulator consists of a current-limiting resistor~$R_{\text{SR}}$, a TL431 linear shunt regulator, and a smoothing capacitance $C_{\text{SR}}$. The MCU must have an analog-to-digital converter (ADC) with at least two channels, as well as at least two general-purpose input/output (GPIO) pins. An MCU was preferred to integrated Qi receivers to allow for more granular control and tuning of WPT parameters during prototype debugging. In the current design, we use an Atmel ATTiny13A-SSU AVR microcontroller. One of the ADC inputs \circlabel{ADC0} is connected to the output of the resistive divider $R_{\text{0H}}$--$R_{\text{0L}}$, which follows the voltage $V_{\text{R}}$ at the rectifier output. To improve noise performance, a filtering capacitance $C_{\text{0L}}$ is connected to the lower leg of the resistor divider.

The circuit has two separate ground nets: the reference ground (denoted as the Earth ground symbol in Fig.~\ref{fig:CircuitDiagram}) which is used as the common wire by the rectifier, the MCU, and the 2.5~V linear regulator; and the power ground (denoted as the chassis ground symbol) used by the rest of the circuit and connected to the negative terminal of the battery. The power ground is connected to the reference ground through a current sense resistor $R_{\text{S}}$. The second ADC input \circlabel{ADC1} of the MCU monitors the voltage across the current-sense resistor with respect to the reference ground. A filtering capacitance $C_{\text{S}}$ is connected between the ADC input and the reference ground to improve noise performance.

%______________________Table_1_______________________
\begin{table}[tbp]
   \centering
   \scriptsize
   \caption{Component values}
   \begin{tabular}{c c c}
\hline\hline
Symbol & Description & Value \\\hline
$L_0$ & Receiving coil inductance & 12 \textmu H \\
$C_0$ & Receiver resonant circuit capacitance & $4\times 68$~nF \\
$C_{\text{M}}$ & Load modulation capacitance & 22 nF \\
$C_{\text{P}}$ & Second resonance capacitance & 2.2 nF \\
$C_{\text{R}}$ & $V_{\text{R}}$~output smoothing capacitance & 20 \textmu F \\
$R_{\text{SR}}$ & Shunt regulator series resistor & 470~\textohm \\
$C_{\text{SR}}$ & MCU supply filter capacitance & 20 \textmu F \\
$R_{\text{0H}}$ & $V_{\text{R}}$ feedback divider upper leg resistance & 16~k\textohm \\
$R_{\text{0L}}$ & $V_{\text{R}}$ feedback divider lower leg resistance & 3~k\textohm \\
$C_{\text{0L}}$ & $V_{\text{R}}$ feedback divider filter capacitance & 0.1~\textmu F \\
$R_{\text{S}}$ & Current sense resistance & 0.1~\textohm \\
$C_{\text{S}}$ & $V_{\text{R}}$ Current sense filter capacitance & 0.1~\textmu F \\
$R_{\text{PD1}}$ & Qi modulation pull-down resistance & 100~k\textohm \\
$R_{\text{PD2}}$ & Power Good pull-down resistance & 10~k\textohm \\
$L_1$ & 5~V buck converter inductance & 4.7~\textmu H 1~A \\
$C_{\text{BST}}$ & 5~V buck converter bootstrap capacitance & 0.1~\textmu F \\
$C_{\text{1H}}$ & 5~V buck converter feed-forward capacitance & 47~pF \\
$R_{\text{1H}}$ & 5~V feedback divider upper leg resistance & 100~k\textohm \\
$R_{\text{1L}}$ & 5~V feedback divider lower leg resistance & 13~k\textohm \\
$C_{\text{F1}}$ & 5~V output smoothing capacitance & 40 \textmu F \\
$C_{\text{CH}}$ & Charger filter capacitance & 47 \textmu F \\
$R_{\text{PROG}}$ & Charge current programming resistance & 10~k\textohm \\
$R_{\text{LED}}$ & LED current limiting resistance & 2.2~k\textohm \\
$C_{\text{BAT}}$ & 1.5~V converter input smoothing capacitance & 47~\textmu F \\
$L_2$ & 1.5~V buck converter inductance & 4.7~\textmu H 1~A \\
$C_{\text{2H}}$ & 1.5~V buck converter feed-forward capacitance & 47~pF \\
$R_{\text{2H}}$ & 1.5~V feedback divider upper leg resistance & 300~k\textohm \\
$R_{\text{2L}}$ & 1.5~V feedback divider lower leg resistance & 200~k\textohm \\
$C_{\text{F2}}$ & 1.5~V output smoothing capacitance & 47 \textmu F \\
\hline
   \end{tabular}
   \label{tab:ComponentValues}
\end{table}
%______________________Table_3_______________________

%______________________Figure_4_______________________
\begin{figure}[tb]
   \centering
   \includegraphics[width=7.5cm]{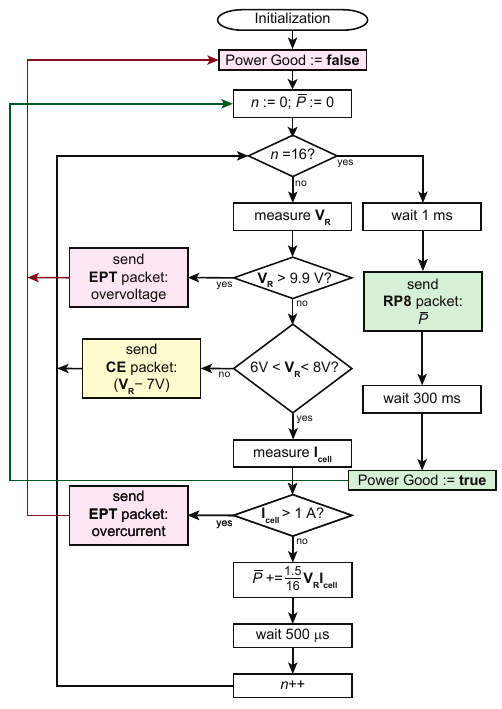}
   \caption{Algorithm of the microcontroller unit (MCU) firmware enabling (when \texttt{Power Good} is assigned \texttt{true}) or disabling (when assigned \texttt{false}) the 5~V buck converter and driving the load modulation to send Control Error (\qipacket{CE}), End of Power Transfer (\qipacket{EPT}) and 8-bit Received Power (\qipacket{RP8}) Qi packets, accomplishing the voltage feedback and foreign object detection. The Signal Strength (\qipacket{SIG}), Identification (\qipacket{ID}) and Configuration (\qipacket{CFG}) packets are sent during the initialization procedure.}
   \label{fig:FirmwareFlowchart}
\end{figure}
%______________________Figure_4_______________________

MCU GPIO outputs provide two logical signals: ``Qi modulation'' (\circlabel{GPIO0}), used to drive the gates of the load modulation MOSFET switches, and ``Power good'' (\circlabel{GPIO1}), used to enable or disable the 5~V buck converter. Both wires are pulled to the reference ground with pull-down resistors $R_{\text{PD1}}$ and $R_{\text{PD2}}$ to ensure that the circuit is idle until the MCU is initialized. The MCU firmware implements an algorithm illustrated by the flow diagram in Fig.~\ref{fig:FirmwareFlowchart}. After initialization, which includes sending the Signal Strength (\qipacket{SIG}), Identification (\qipacket{ID}) and Configuration (\qipacket{CFG}) packets to the charging station to establish the power transfer contract according to the Qi~1.1 Baseline protocol~\cite{QiStandard}, the MCU starts to perform periodic measurements of the rectifier output voltage $V_{\text{R}}$ and the current $I_{\text{cell}}$ consumed by the battery. If an overvoltage or an overcurrent event is detected, an End of Power Transfer (\qipacket{EPT}) packet is sent to the transmitter to terminate the power transfer contract immediately. If the voltage is found to be out of the desired range ($6~\text{V} < V_{\text{R}} < 8~\text{V}$), a Control Error (\qipacket{CE}) packet containing the deviation $\Delta V = (V_{\text{R}} - 7~\text{V})$ in arbitrary units (converted to a signed 8-bit integer) is sent to the transmitter, requesting it to increase or decrease the transmitted power. To measure the received power, the MCU performs voltage $V_{\text{R}}$ and current $I_{\text{cell}}$ measurements and calculates the received power as
\begin{equation}
    P_{n} = 1.5 V_{\text{R}}I_{\text{cell}},
\end{equation}
where the factor of 1.5 has been introduced to account for AC power absorption~\cite{QiStandard}. The instantaneous values of the received power $P_n$ are obtained once every 500~\textmu s. Finally, an averaged value $\bar{P}$
\begin{equation}
    \bar{P} = \frac{255}{5~\text{W}}\cdot\frac{1}{16}\sum\limits_{n = 1}^{16} P_n
\end{equation}
is calculated over 16 samples, converted to an unsigned 8-bit integer and sent as the 8-bit Received Power (\qipacket{RP8}) packet to the transmitter.

Fig.~\ref{fig:CircuitDiagram}~(c) shows the implementation of the 5~V buck converter. The driver chip must be an adjustable 600~kHz current-mode converter with a 0.6~V reference voltage (\circlabel{FB}), comprising a bootstrapping circuit (\circlabel{BST}) and a logical enable signal (\circlabel{EN}). We utilized an LP6460A converter for the current design. The buck converter uses the inductance $L_1$ to store energy, the capacitor $C_{\text{BST}}$ for high-side bootstrapping, as well as the capacitor $C_{\text{F1}}$ to smooth out the switching noise. The feedback circuit consists of two resistors $R_{\text{1H}}$ and $R_{\text{1L}}$, as well as a feed-forward capacitor $C_{\text{1H}}$.

The circuit implementation of the converter board, which comprises the Li-ion battery charger and the 1.5~V buck converter, is shown in Fig.~\ref{fig:CircuitDiagram}. The battery charge circuit uses an analog of the LTC4054-4.2 standalone charger, which comprises an open collector output (\circlabel[\overline]{CHRG}) indicating the charging state and a current programming input (\circlabel{PROG}) for setting the current used in the constant current charge mode. Despite this charger being a dissipative one, it was chosen due to its compact size and availability. In the current design, we used the STC4054 chip. The charge current is programmed to $I_{\text{CC}} = 100~\text{mA}$ by the resistor $R_{\text{PROG}}$. A light emitting diode (LED) with resistor $R_{\text{LED}}$ is added for a visible indication of the charge state. The 1.5~V buck converter is implemented using an adjustable 1.7~MHz converter with current limiting and a 0.6~V reference voltage (\circlabel{FB}). We used a NCP1529ASNT1G converter for the current design. The buck converter uses the inductance $L_2$ and the smoothing capacitor $C_{\text{F2}}$ to store energy, while the resistors $R_{\text{2H}}$ and $R_{\text{2L}}$, as well as the feed-forward capacitor $C_{\text{2H}}$ comprise the feedback circuit.

%____________________Numerical_Simulations____________________
\section{Numerical simulations of magnetic coupling between transmitting and receiving coils}
\label{sec:Simulations}

To address the feasibility of the proposed concept, we proceed with numerical simulations and experimental studies of curved receiving coils. In particular, we consider the effects of the coil curvature radius, its spatial separation from the transmitting coil, and its rotation along the battery axis on the WPT efficiency and magnetic field distributions.

%______________________Figure_5_______________________
\begin{figure*}[tbp]
   \centering
   \includegraphics[width=14cm]{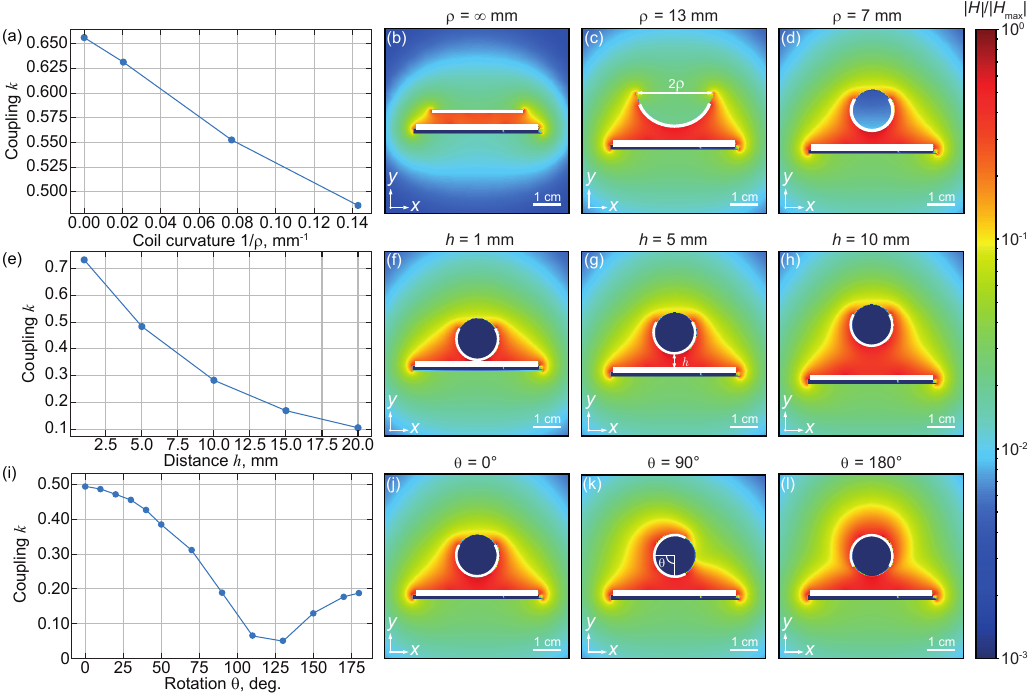}
   \caption{(a) Numerically calculated coupling coefficient $k$ between the coils at $f=100$~kHz for different values of curvature radius $\rho$ of the receiving coil: $\rho=\infty$ for a planar coil (blue), $\rho=49$~mm (red), $\rho=13$~mm (green), and $\rho=7$~mm corresponding to an AA battery (orange). (b)-(d) Magnetic field profiles in the (xy)-plane passing through the battery midpoint for the receiving coils with curvature radii (b) $\rho = \infty$, (c) $\rho = 13$~mm, and (d) $\rho = 7$~mm, respectively. (e) Coupling between the coils at $f=100$~kHz for different distances $h$ between a transmitting coil and a curved receiving coil ($\rho = 7$~mm). (f)-(h) Magnetic field profiles in the (xy)-plane for (f) $h=1$~mm, (g) $h=5$~mm, and (h) $h=10$~mm. (i) Coupling between the coils at $f=100$~kHz for different receiving coil rotation angles $\theta$; $\rho = 7$~mm. (j)-(l) Magnetic field profiles in the (xy)-plane for (j) $\theta=0^{\circ}$, (k) $\theta=90^{\circ}$, and (l) $\theta=180^{\circ}$.}
   \label{fig:Simulations}
\end{figure*}
%______________________Figure_5_______________________

%______________________Figure_6_______________________
\begin{figure}[t!]
   \centering
   \includegraphics[width=8.5cm]{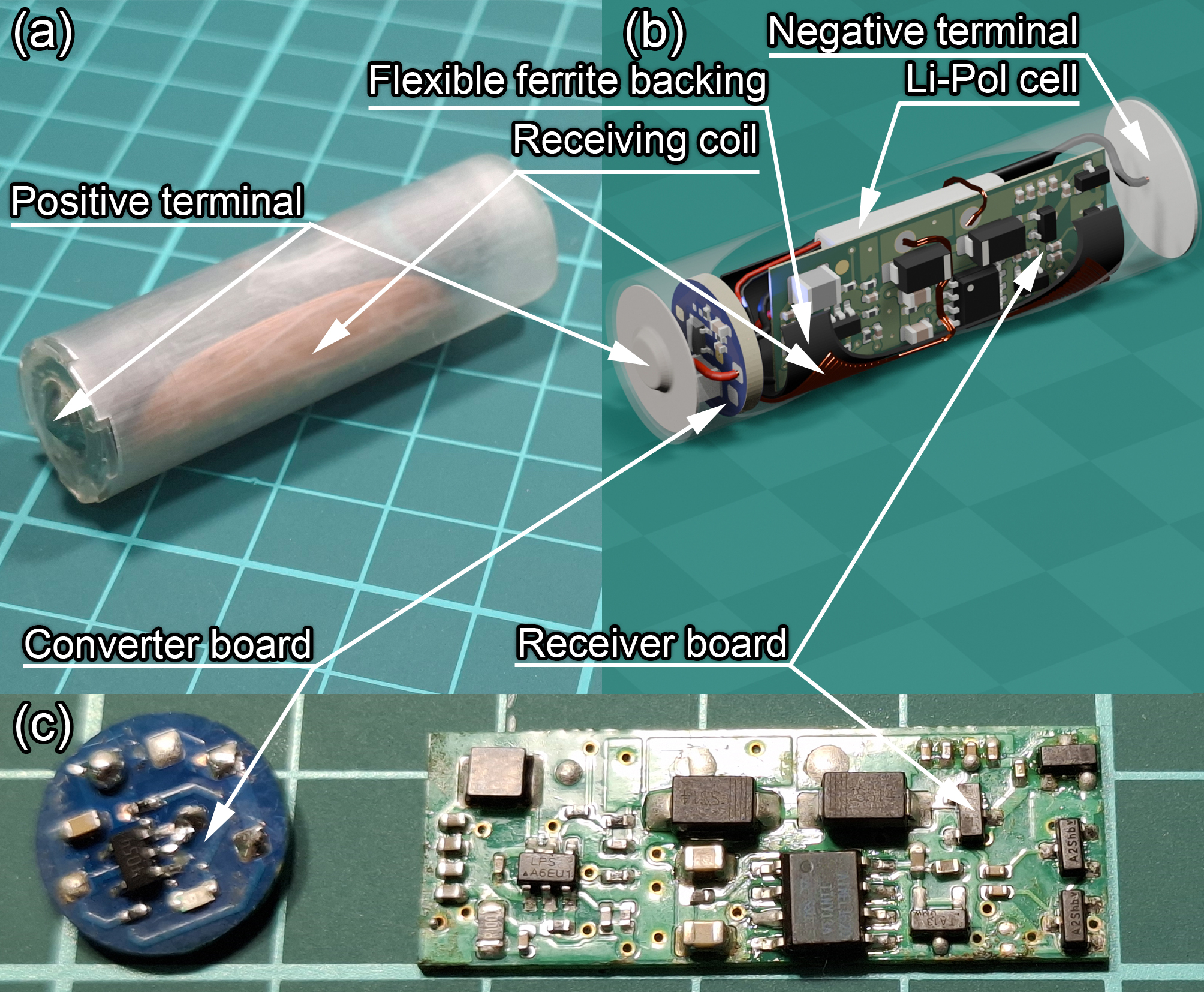}
   \caption{Photograph of the battery prototype (a) and a sectional diagram of the same battery (b) showing the receiving coil with a flexible ferrite backing sheet, the Li-ion cell, the receiver board, the converter board, as well as positive and negative terminals. (c) Photograph of the assembled receiver and converter boards.}
   \label{fig:Prototype}
\end{figure}
%______________________Figure_6_______________________

%______________________Figure_7_______________________
\begin{figure*}[tp]
   \centering
   \includegraphics[width=14cm]{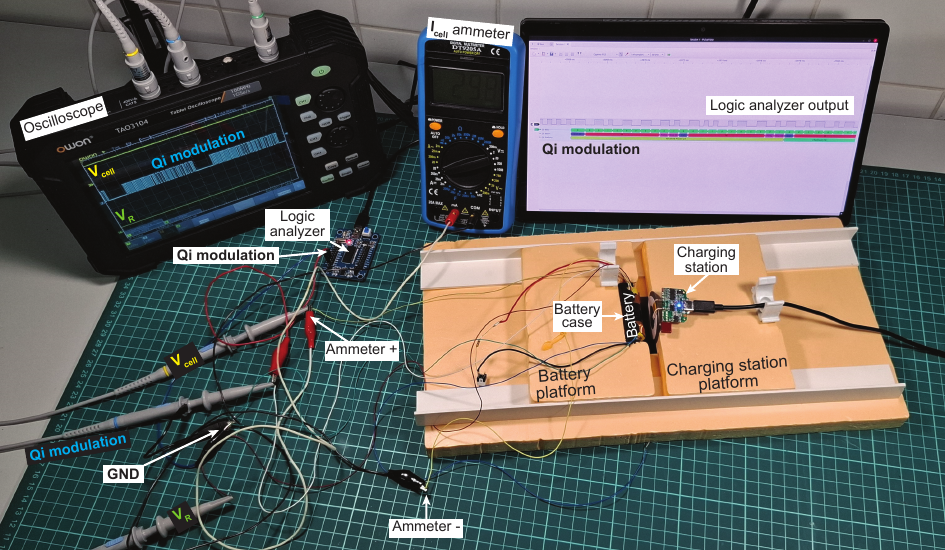}
   \caption{Photograph of the experimental setup for measurement of the distances between the transmitting and receiving coils at which a power transfer contract is established and terminated at various battery orientation angles, as well as for charge curve measurement.}
   \label{fig:ExperimentalSetup}
\end{figure*}
%______________________Figure_7_______________________

%______________________Figure_8_______________________
\begin{figure}[t!]
   \centering
   \includegraphics{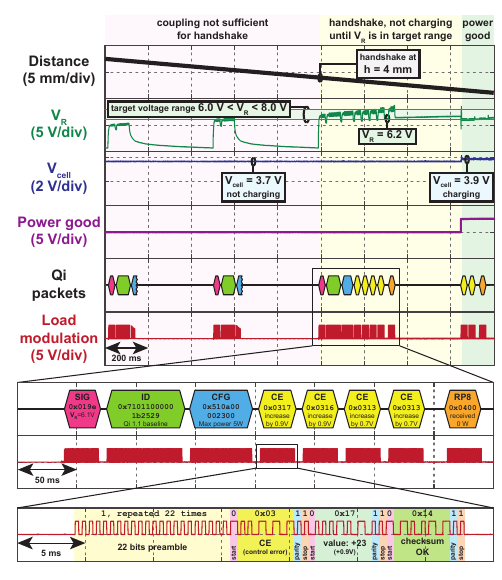}
   \caption{Timing diagram demonstrating the establishment of a power transfer contract (abbreviated as ``handshake'' in the figure). Top panel shows experimentally measured voltage at the rectifier output ($V_{\text{R}}$), Li-ion cell voltage ($V_{\text{cell}}$), ``Power Good'' signal voltage (2.5 V logic), ``Load modulation'' signal voltage (2.5 V logic) as well as decoded Qi packets, while the distance between the battery and the charging station is varied from 7 to 2~mm with the power transfer contract establishment occurring at 4~mm. Middle panel shows a number of decoded Qi packets with a magnified time scale, namely a Signal Strength (SIG) packet, an Identification (ID) packet, a Configuration (CFG) packet, four Control Error (CE) packets as well as a 8-bit Received Power (RP8) packet. Bottom panel shows the load modulation signal comprising a single Control Error packet in a further magnified time scale.}
   \label{fig:Oscillograms}
\end{figure}
%______________________Figure_8_______________________

We perform all numerical simulations with CST Microwave Studio applying the finite element method (FEM) solver in the frequency domain. The model includes a receiving coil and a transmitting coil supplied with ferrite backing layers, as well as two ports: Port~$1$ for the transmitter and Port~$2$ for the receiver, correspondingly. Inside the outline of the AA battery that is covered by the receiving coil, a layered cylinder with length $40$~mm and diameter $13.4$~mm is placed to emulate the electromagnetic properties of a Li-ion cell and the required electronics. 

%______________________Coil_curvature_______________________
The outline of an AA battery is a cylinder $49.5..50.5$~mm in length with $13.7..14.5$~mm diameter. Since the characteristic in-plane dimensions of a typical Qi receiving coil are $30 \times 40$~mm, incorporating such a coil inside an AA battery requires its bending. In turn, bending the coil changes the magnetic field distribution and the resonant frequency of the receiver~\cite{2020_Wen_Curvature,2023_Zhuang_Omnidirectional}. We start by examining how the radius of curvature $\rho$ of the receiving coil bent around a cylindrical surface in the $(xy)$-plane in Fig.~\ref{fig:Simulations} affects the coupling between the receiving and transmitting coils. To proceed, we perform numerical simulations for the models with a flat receiving coil ($\rho = \infty$), as well as for the receiving coils with curvature radii $\rho = 49$~mm, $\rho = 13$~mm, and $\rho = 7$~mm, Fig.~\ref{fig:Simulations}(a)-(d). The distance between the receiving and transmitting coils in all cases is $h=5$~mm, and the receiving coil is supplied with a thin ferrite layer. As seen in Fig.~\ref{fig:Simulations}(a), the coupling coefficient $k = M/\sqrt{L_1L_2}$ between the coils only decreases by 25\% after bending the coil to a curvature radius $\rho = 7$~mm.

The absolute values of the magnetic field amplitude in the $(xy)$-plane $H(x,y)$ are shown in Fig.~\ref{fig:Simulations}(b)-(d) for $\rho = \infty$, $\rho = 13$~mm, and $\rho = 7$~mm, respectively. The magnetic field concentrates between the transmitting coil and the planar receiving coil homogeneously in Fig.~\ref{fig:Simulations}(b), while forming a hot spot near the central region in Fig.~\ref{fig:Simulations}(c) corresponding to the closest distance between the transmitting coil and the curved receiving coil with $\rho=13$~mm. Although the receiving coil is covered with a ferrite layer from the inner surface, the field is still present in the inner region of a cylindrical battery outline, which may lead to undesired couplings with other components if they are placed in the respective region. Finally, for the receiving coil with $\rho=7$~mm corresponding to the AA battery size, the magnetic field hot spot becomes more pronounced, while the field no longer penetrates into the inner region, Fig.~\ref{fig:Simulations}(d). For both curved receiving coil geometries considered in Fig.~\ref{fig:Simulations}(c,d), the magnetic field distribution in the surrounding free space changes insignificantly, indicating that there should not arise any increased couplings with the environment or undesired electromagnetic fields acting on the user compared to the commercially available planar Qi receivers.

%______________________Coil_distance_______________________
Next, we consider the coupling and magnetic field distributions on the distance between the transmitting coil and the receiving coil with a curvature radius $\rho=7$~mm. As seen in Fig.~\ref{fig:Simulations}(e), the dependence of coupling $k$ on the distance $h$ between the transmitting coil and the protruding part of the receiving coil demonstrates a hyperbolic decrease, similarly to the case of planar coils. However, couplings $k \ge 0.3$ are achieved for distances up to 10~mm, which appears adequate for power transfer with the Qi protocol. Magnetic field distributions shown in Fig.~\ref{fig:Simulations}(f)-(h) demonstrate a vertical elongation of the field concentration region between the nearest points of the transmitting and receiving coils upon increasing the separation $h$ from $1$~mm to $10$~mm, with a constriction in the field distribution formed for $h=10$~mm in Fig.~\ref{fig:Simulations}(h) caused by the field localization near the coils' surfaces.

%______________________Coil_rotation_______________________
Finally, we examine the influence of the rotation of the receiving coil. Figure~\ref{fig:Simulations}(i) demonstrates couplings for rotation angles from $\theta=0^{\circ}$ to $\theta=180^{\circ}$. The system with the receiving coil rotated by $\theta=180^{\circ}$ demonstrates a considerably coupling compared to the basic configuration with $\rho=7$~mm, $h=5$~mm, and $\theta=0^{\circ}$. However, the lowest coupling is observed for $\theta=130^{\circ}$, as seen in Fig.~\ref{fig:Simulations}(i). Qualitatively, a considerable part of the receiving coil surface becomes orthogonal to the transmitting coil plane, and thus parallel to the magnetic field direction, which results in lower coupling between the coils. In contrast to previous cases, the magnetic field profiles for $\theta=90^{\circ}$ demonstrate a pronounced asymmetry in the vicinity of the receiving coil; see Fig.~\ref{fig:Simulations}(k). Finally, we note that $k \ge 0.3$ for the rotation angles $\theta\le60^{\circ}$. The coupling $k \simeq 0.3$ is the minimum value allowing establishment of a power transfer contract (experimental measurements of coupling are described in the Supplementary material~\cite{Supplementary}), therefore, sufficient coupling is expected unless the rotation angle exceeds $60^{\circ}$.

%________________________Experiments__________________________
\section{Experimental studies of the battery prototype}
\label{sec:Experiments}

To demonstrate the practical applicability of our concept, we assemble a battery prototype shown in Fig.~\ref{fig:Prototype} inside a casing having the shape of an empty cylinder with the height of $47.8$~mm, the diameter of $14$~mm, and the wall thickness of $0.5$~mm, which was manufactured from photopolymer resin at the Anycubic Photon SLA 3D printer. We study its ability to establish a Qi power transfer contract using load modulation and to receive enough power to charge the Li-ion cell at a 100~mA current at various distances from the charge station and rotation angles around the battery axis. Fig.~\ref{fig:ExperimentalSetup} shows the experimental setup used for these experiments, which includes: (i) a platform (Battery platform) with an AA battery case (Battery case), inside which the battery (Battery) prototype is installed in a way allowing rotation around its axis; (ii) a platform (Charging station platform) with a Qi charging station (Charging station) which has been taken out from the casing to enable precise measurement of the distance between the transmitting and receiving coils; (iii) an ammeter ($I_{\text{cell}}$ ammeter) showing the cell charge current $I_{\text{cell}}$; (iv) an Owon TAO3104 oscilloscope (Oscilloscope) used for measurement of the rectifier output voltage $V_{\text{R}}$, the Li-ion cell voltage $V_{\text{cell}}$ and the 2.5~V logical level signal at the gates of load modulation MOSFET switches (Qi modulation); (v) an fx2lafw~\cite{Fx2lafw} compatible logic analyzer (Logic analyzer) connected to the 2.5~V logical level signal at the gates of load modulation MOSFET switches (Qi modulation) and (vi) a personal computer showing the decoded Qi packets using the Sigrok PulseView software (Logic analyzer output). The battery platform and the charging station platform are made of extruded polystyrene foam to ensure the absence of coupling and are movable in a single direction to facilitate the variation of the distance between the battery and the charging station.
 
We start by testing the establishment of a power transfer contract between the battery and the charging station. For that purpose, the rectifier output voltage $V_{\text{R}}$, the Li-ion cell voltage $V_{\text{cell}}$ and the 2.5~V logical level signal driving the gates of load modulation MOSFETs (Qi modulation) as well as the 2.5~V logical level signal enabling the 5~V buck converter (Power good) were recoded into the oscilloscope memory while the charging station was pushed toward the battery prototype until power transfer was initiated. Fig.~\ref{fig:Oscillograms} shows the obtained signals, as well as the decoded Qi packets in the form of a timing diagram. First, at distances greater than 4~mm, two digital ping pulses with a 100~ms width sent by the charging station can be observed. However, the coupling between the transmitting and receiving resonant circuits is not strong enough at this point for the capacitive load modulation to invoke a significant change in the current consumed by the transmitter (the 5~V buck converter is disabled at this point, while the consumption current of the MCU does not exceed 100~\textmu A). Therefore, the charge station is unable to ensure the presence of a Qi-compatible receiver at these distances and does not start transmitting power continuously. Instead, it ends each ping pulse after 100~ms, while the battery is in the middle of sending a Configuration (\qipacket{CFG}) packet.

Once the battery is within 4 mm of the charging station, the coupling becomes strong enough for the load modulation to cause changes in the transmitter's current consumption large enough to be measured and decoded. Therefore, the charger station successfully receives the Signal Strength (\qipacket{SIG}) and Identification (\qipacket{ID}) packets sent by the battery and switches to continuous power transmission, so the Configuration (\qipacket{CFG}) packet is also successfully transferred. The \qipacket{ID} packet is used to choose the Qi~v~1.1 Baseline protocol as well as to identify the receiver to the station using a Manufacturer Code (in the current design, a bogus manufacturer code \texttt{0x1000} was used) and a basic device identifier which can be generated dynamically (we used a number \texttt{0x001b2529} in the current design)~\cite{QiStandard}. The \qipacket{CFG} packet tunes the charge station to the following settings: (i) Extended protocol not supported, no optional configuration packets; (ii) 5~W reference power; (iii) 8~ms power measurement window with a 12~ms offset; (iv) authentication and out-of-band communications functionalities not supported. The last byte of the packet, which contains the parameters related to bidirectional communication used in the extended protocol, was set to zero. After the \qipacket{CFG} packet, the battery sends a number of Control Error (\qipacket{CE}) packets containing the deviation of the rectifier output voltage $V_{\text{R}}$ from the target level of 7~V, converted to a 8-bit signed integer in arbitrary units. It stops sending the \qipacket{CE} packets as soon as $6~\text{V} < V_{\text{R}} < 8~\text{V}$. Finally, after the power measurement window ends, the battery sends an 8-bit Received Power (\qipacket{RP8}) packet containing the received power averaged over 16 samples measured each 500~\textmu s during the 8-ms window, converted to an 8-bit unsigned integer.

%______________________Figure_9_______________________
\begin{figure}[tb]
   \centering
   \includegraphics{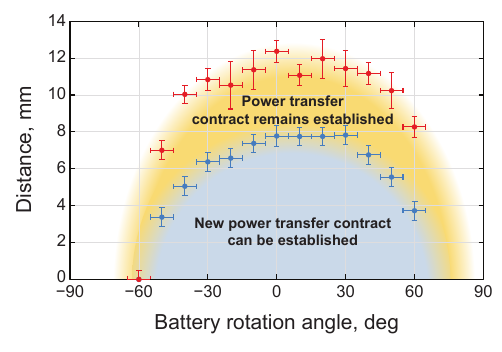}
   \caption{Experimentally measured distances between the transmitting and receiving coils, which correspond to power transfer contract establishment (blue markers) and termination (red markers) at various battery orientation angles. The blue and orange colored areas are guides to the eye showing the ranges of battery rotation angles and distances between the transmitting and receiving coils allowing power transfer contract establishment (blue area) and battery charging using an already established power transfer contract (orange area).}
   \label{fig:ActionRadius}
\end{figure}
%______________________Figure_9_______________________

After a 300-millisecond pause from the first RP8 packet, the 5~V buck converter is enabled by setting the Power good signal to logical one. This is accompanied by an increase of the cell voltage from $V_{\text{cell}} = 3.7~\text{V}$ to $V_{\text{cell}} = 3.9~\text{V}$ as well as a decrease of the rectifier output voltage $V_{\text{R}}$ to 5.9~V, which now falls out of the desired range. Therefore, the battery sends two \qipacket{CE} packets requesting the charge station to increase the power. As soon as $V_{\text{R}}$ becomes greater than $0.6~\text{V}$, another \qipacket{RP8} packet is sent.

%______________________Table_2_______________________
\begin{table*}[t]
   \centering
   \scriptsize
   \caption{Comparative analysis of batteries with a wireless charging functionality and wireless power transfer systems incorporating curved transmitting or receiving coils}

   \begin{tabular*}{\textwidth}{@{\extracolsep{\fill}} c c c c c c c c}
   \hline\hline
   \multicolumn{8}{c}{\textsc{Batteries with wireless charging}} \\
   \hline
   Reference & WPT technology & WPT frequency & \parbox[c][8mm]{1.2cm}{\centering max WPT\\Rx power} & Power storage & Energy capacity & \parbox[c][8mm]{1.2cm}{\centering Charging\\duration} & Size  \\
   \hline
   \cite{2021_Smith} & electrodynamic & 238~Hz & 240~\textmu W & GM300910 Li-Pol cell & 44 mW$\cdot$h & n/a & AA \\
   \cite{2016_Bidul} & magnetic resonant (Qi) & 100--200~kHz & 5~W & Li-Pol cell & 7400 mW$\cdot$h & 1~H & custom \\
   \cite{Patent_CN106340940B} & magnetic resonant (Qi) & 100--200~kHz & 5~W & NiMH cell & 1560 mW$\cdot$h & 2~H & n/a \\
   \cite{2023_Zens,Patent_NL2021406B1} & magnetic resonant (Qi) & 100-200~kHz & n/a & Li-Pol cell & 14800~mW$\cdot$h & $\sim 4~\text{H}$ & $96\kern-2pt\times\kern-2pt{}64\kern-2pt\times\kern-2pt{}17~\text{mm}^3$ \\
   %power bank \\
   This work & magnetic resonant (Qi) & 100--200~kHz & 5~W & LP451124 Li-Pol cell & 240 mW$\cdot$h & 1.5~H & AA \\
   \hline
   \end{tabular*}
   
   \vspace{1.5pt}
   
   \begin{tabular*}{\textwidth}{@{\extracolsep{\fill}} c c c c c c c}
   \hline
   \multicolumn{7}{c}{\textsc{Wireless power transfer using curved coils}} \\
   \hline
   Reference & WPT technology & WPT frequency & WPT power & Bent coil & Coil size & Curvature radius \\\hline
   \cite{2020_Wen_Curvature} & magnetic resonant & 200~kHz & n/a & receiving & $\varnothing 18~\text{cm}$ &  $80\ldots200~\text{mm}$ \\
   \cite{2023_Zhuang_Omnidirectional} & magnetic resonant & 218~kHz & 30~W & transmitting & $300\times148~\text{mm}^2$ & 200~mm \\
   \cite{2019_Sondhi} & electrodynamic & $<1~\text{kHz}$ & $5\ldots36~\text{W}$ & transmitting & $\varnothing 10~\text{cm}$ & $16~\text{mm}\ldots\infty$
   \end{tabular*}
   \label{tab:ComparisonWithAnalogs}
\end{table*}
%______________________Table_2_______________________

We proceed with evaluating the distances between the battery and the charging station at which a power transfer contract can be established for various battery rotation angles. Specifically, we first set the battery at the given orientation angle around its axis and move the charging station platform 50~mm away from the battery. Then the charging station is slowly pushed toward the battery until a power transfer contract is established, which is identified by the appearance of Qi \qipacket{CE} packets, as well as by a change in the LED indication pattern of the charging station. Then the charging station platform is pulled back away from the battery with the same speed, until the power transfer contract terminates due to weak coupling, which is identified by a change in LED indication of the charging station. The measured distance was then averaged by 10 measurements. Fig.~\ref{fig:ActionRadius} shows these averaged distances with markers. The horizontal error bars correspond to a $\pm 5^{\circ}$ measurement error of the rotation angle, while the vertical bars account for a measurement error of 0.5~mm as well as the standard deviation between the 10 measurements of the distance. Significant hysteresis is noticeable from the experimental data: the distance of power transfer contract termination is almost twice the distance of its establishment. This can be explained by the fact that the battery only consumes negligible current (less than 100~\textmu A) at the time of the establishment of a power transfer contract, but after it is established, the drawn current increases by three orders of magnitude, allowing for a deeper load modulation. Therefore, changes in the current consumed by the transmitter that are large enough to be measured occur at larger distances between the charge station and the battery.

\begin{figure}[t]
   \centering
   \includegraphics{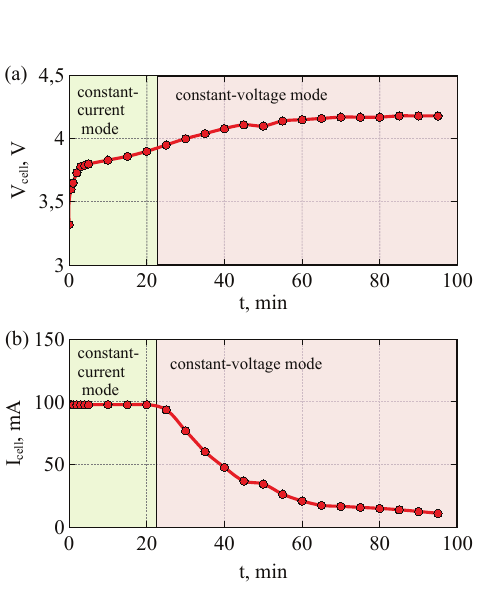}
   \caption{Measured time dependence of the Li-ion cell voltage $V_{\text{cell}}$ and charge current $I_{\text{cell}}$ for the prototype receiving power from a Qi standard charging station.}
   \label{fig:ChargeCurve}
\end{figure}

To evaluate the time needed to fully charge the battery using WPT, we obtained its charge curves with the assembled battery placed 2~mm away from the transmitting coil of a Qi compatible charging station. An ammeter was plugged into the positive cable of the Li-ion cell, while the voltage $V_{\text{cell}}$ across the positive and negative leads of the cell was measured using the oscilloscope. The measured curves are shown with solid lines in Fig.~\ref{fig:ChargeCurve}. The first $22$~minutes correspond to the constant current (CC) charging mode at $I_{\text{CC}} = 100~\text{mA}$, which is characterized by a rapid voltage increase. After that, the STC4054 chip made a transition to the constant voltage (CV) mode, which is characterized by a gradual decrease in charging current accompanied by a slow increase in $V_{\text{cell}}$. After $95$~minutes of charging, $I_{\text{cell}}$ falls below the cut-off current, defined as $0.1I_{\text{CC}}$, at which point the charging is considered complete and the STC4054 chip disconnects the cell from the charging source. 

%______________________Figure_10_______________________
%______________________Figure_10_______________________

Finally, we evaluate the efficiency of the wireless charging. The detailed procedure and results are described in the Supplementary Material~\cite{Supplementary}. The efficiency decreases with the distance $h$ between the transmitting and receiving coils, as well as the state of charge (SoC) of the battery, with the highest value being 28\% for $h = 0~\text{mm}$ and $\text{SoC} = 0\%$. \blue{The most prominent loss channels are WPT losses with $\eta_{\text{AC-AC}}(h, \text{SoC}) \le 50\%$, DC-DC converter losses with $\eta_{\text{DC-DC}}(\text{SoC})$ in the range of 40\%--80\%, as well as STC4054 charger losses with $\eta_{4054}(\text{SoC})$ in the range of 66\%--83\%. Low values of $\eta_{\text{AC-AC}}$ correspond to the coupling $k < 0.75$ between the transmitting and the bent receiving coils. Therefore, the main strategy for efficiency improvement would be the replacement of the DC-DC converter and the STC4054 charge controller by a specialized integrated circuit that would utilize the Qi feedback loop to implement CC-mode and CV-mode charging control by regulation of the transmitted power.}

%____________________Disscussion______________________
\section{Conclusion and Outlook}
\label{sec:Outlook}

To conclude, we have proposed the concept of a wirelessly charged battery with the size and output voltage of a standard AA battery, supplied with a curved Qi receiving coil. Examining the proposed concept numerically and experimentally, we demonstrate that the application of a receiving coil bent around the inner side of a cylindrical battery casing provides sufficient coupling to implement wireless charging and demonstrates WPT efficiency close to that of a standard receiver with a planar coil. We also study the effects of battery rotation around its symmetry axis and spatial separation between the battery and the transmitting coil and demonstrate that a power transfer contract can be established for separations up to 8 mm when the coils are aligned, and rotations up to $60^{\circ}$ from the direction of coil alignment when there is no separation between the coils. Finally, we validate the concept experimentally by constructing a fully operational prototype. Compared to other proposed batteries with a wireless charging capability, our prototype offers a compromise in energy capacity and charging time, while being beneficial by simultaneous compatibility with AA battery holders and Qi-compliant charging stations; see Table~\ref{tab:ComparisonWithAnalogs} for a detailed comparison.

The directions of further development are the following. \blue{First, the proof-of-concept design presented in this work can be enhanced from an engineering point of view by optimizing the receiving coil~\cite{2016_Cove_Improving,2019_Li_Single,2021_Rituraj_New}, making the receiver and converter circuits more efficient and compact by integrating them on an application-specific integrated circuit, thus reducing the number of power conversion steps and providing more space for a rechargeable cell, as well as by usage of batteries with a higher power density, such as high-voltage Li-ion cells~\cite{2023_Frith}.} Second, the design can be improved by incorporating solutions to increase the transmission range while maintaining compatibility with the Qi standard~\cite{2020_Sousa_Optimal}. Furthermore, interactions between several receivers should be considered~\cite{2021_Wu_Node, 2025_Zolotarev_Multi}, corresponding in our case to powering up a certain device with several wirelessly charged batteries simultaneously. Another direction of improvement is the introduction of simultaneous wireless power and data transfer capability~\cite{2024_Jing_Simultaneous,2026_Luo_Cost}. Finally, several WPT concepts and standards are under active development, including omnidirectional~\cite{2020_Kim_Plane} and room-scale volumetric wireless charging working at frequencies from $1$~MHz~\cite{2024_Mikhailov} to $1.34$~MHz~\cite{2021_Sasatani}. Developing a wirelessly charged battery compatible simultaneously with any of the mentioned standards and Qi renders feasible, see the examples of a composite Qi-AirFuel transmitter~\cite{2024_Moisello} and Qi-compatible multi-mode receivers~\cite{2021_Oh_15,2023_Shah_Design}, yet very interesting task, capable of bringing universal wireless charging solutions.

%____________________Acknowledgments____________________
\section*{Acknowledgment}
We acknowledge fruitful discussions with Eugene Koreshin, Aigerim Jandaliyeva, Polina Kapitanova, and Alexey Slobozhanyuk. The work was supported by grant No.~FSER-2024-0041 within the framework of the national project "Science and Universities".

\end{document}

%\preprint{AIP/123-QED}

\title{Supplemental Material\\A rechargeable AA battery supporting Qi wireless charging}

\author{Alexey~A.~Dmitriev}
\affiliation{School of Physics and Engineering, ITMO University, 49 Kronverksky pr., bldg. A, 197101 Saint Petersburg, Russia}

\author{Egor~D.~Demeshko}
\affiliation{School of Physics and Engineering, ITMO University, 49 Kronverksky pr., bldg. A, 197101 Saint Petersburg, Russia}

\author{Danil~A.~Chernomorov}
\affiliation{School of Physics and Engineering, ITMO University, 49 Kronverksky pr., bldg. A, 197101 Saint Petersburg, Russia}

\author{Andrei~A.~Mineev}
\affiliation{School of Physics and Engineering, ITMO University, 49 Kronverksky pr., bldg. A, 197101 Saint Petersburg, Russia}

\author{Oleg~I.~Burmistrov}
\affiliation{School of Physics and Engineering, ITMO University, 49 Kronverksky pr., bldg. A, 197101 Saint Petersburg, Russia}

\author{Sergey~S.~Ermakov}
\affiliation{School of Physics and Engineering, ITMO University, 49 Kronverksky pr., bldg. A, 197101 Saint Petersburg, Russia}

\author{Alina~D.~Rozenblit}
\affiliation{School of Physics and Engineering, ITMO University, 49 Kronverksky pr., bldg. A, 197101 Saint Petersburg, Russia}

\author{Pavel~S.~Seregin}
\affiliation{School of Physics and Engineering, ITMO University, 49 Kronverksky pr., bldg. A, 197101 Saint Petersburg, Russia}

\author{Nikita~A.~Olekhno}
   \email{nikita.olekhno@metalab.ifmo.ru}
\affiliation{School of Physics and Engineering, ITMO University, 49 Kronverksky pr., bldg. A, 197101 Saint Petersburg, Russia}

\date{\today}

\maketitle

\begin{spacing}{1.5}

\tableofcontents

%______________________S1_Planar_coils_Theory_______________________
\section*{Supplementary Note 1. Evaluation of planar receiving coils}

We start by analyzing whether a planar receiving coil placed inside an AA battery allows achieving a sufficient coupling to a standard Qi transmitting coil for the establishment of a power transfer contract. A standard A11-type transmitting coil commonly used in Qi charging stations has an outer diameter of $2R_{\text{out}} = 42$~mm. Meanwhile, for a planar coil to fit inside the casing of an AA battery, its diameter must not exceed $2r_{\text{out}} = 14$~mm, which is several times smaller than the transmitting coil. Therefore, to estimate the maximal coupling, we can assume that the magnetic field penetrating the receiving coil is uniform. For such an estimate, we also treat the coils as simple current loops. The maximal magnetic field $B$ achieved in the system, which corresponds to the receiving and transmitting coils lying in the same plane and on the same axis, is expressed as 
\begin{equation}
    B = \mu_0\frac{N_{\text{Tx}}I_{\text{Tx}}}{2R_{\text{avg}}},
\end{equation}
as given by the Biot-Savart law, where $N_{\text{Tx}}$ is the number of turns in the transmitting coil, $I_{\text{Tx}}$ is the current passing through this coil and $R_{\text{avg}}$ is its average radius. For a uniform magnetic field, the magnetic flux $\Phi_{\text{Rx}}$ through the receiving coil is
\begin{equation}
    \Phi_{\text{Rx}} = B N_{\text{Rx}}{\pi r_{\text{avg}}^2},
\end{equation}
where $N_{\text{Rx}}$ is the number of turns in the receiving coil and $r_{\text{avg}}$ is its average radius. Therefore, the mutual inductance $M = \Phi_{\text{Rx}} / I_{\text{Tx}}$ evaluates to
\begin{equation}
    M = \mu_0\frac{N_{\text{Tx}}N_{\text{Rx}}}{2R_{\text{avg}}} \pi r_{\text{avg}}^2.
\end{equation}

The self-inductance of a planar spiral coil is given by~\cite{1999_Mohan_Simple}
\begin{equation}
  L = \mu\mu_0N^2 R_{\text{avg}} c_1\left(\ln\frac{c_2}{\rho} +c_3\rho + c_4\rho^2 \right),
\end{equation}
where $N$ is the number of turns, $\rho = (R_{\text{out}} - R_{\text{in}})/(R_{\text{out}} + R_{\text{in}})$, the average radius depends on the outer radius $R_{\text{out}}$ and the inner radius $R_{\text{in}}$ as $R_{\text{avg}} = (R_{\text{out}} + R_{\text{in}})/2$. For a circular coil, the values of other parameters are $c_1 = 1$, $c_2 = 2.46$, $c_3 = 0$ and $c_4 = 0.2$. We assume $R_{\text{out}} \gg R_{\text{in}}$ for both coils in this estimate, therefore $\rho \simeq 1$ and $R_{\text{avg}} \simeq R_{\text{out}}/2$. Then, the self-inductances $L_{\text{Tx}}$ and $L_{\text{Rx}}$ are given by
\begin{align}
  L_{\text{Tx}} &= 1.1\mu_0N_{\text{Tx}}^2 \frac{R_{\text{out}}}{2}, \\
  L_{\text{Rx}} &= 1.1\mu_0N_{\text{Rx}}^2 \frac{r_{\text{out}}}{2}.
\end{align}

Finally, the magnetic coupling $k = M/\sqrt{L_{\text{Tx}}L_{\text{Rx}}}$ takes the form
\begin{equation}
    k \simeq \frac{\pi}{2.2} \left(\frac{r_{\text{out}}}{R_{\text{out}}}\right)^{1.5}.
\end{equation}
For $R_{\text{out}} = 21~\text{mm}$ and $r_{\text{out}} = 7~\text{mm}$, $k \simeq 0.27$, which is lower than the minimum coupling coefficient necessary for the proper establishment of a Qi power transfer contract (see Fig.~\ref{fig:Sparameters}(b) and Fig.~9 of the main text). Therefore, it is not possible to achieve wireless power transfer (WPT) from a standard Qi charging station to a planar receiving coil of a suitable size for an AA battery.

%__________________S2_Numerical_Simulations__________________
\section*{Supplementary Note 2. Numerical simulations}

%______________________Figure_S1_______________________
\begin{singlespace}
\begin{figure*}[b]
    \includegraphics[width=8.5cm]{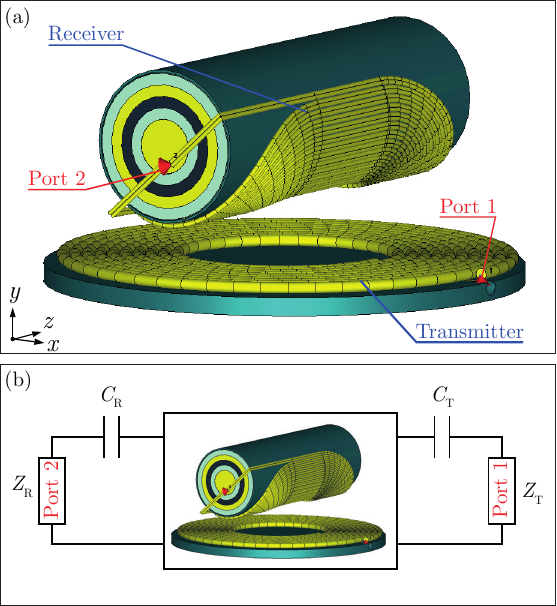}
    \caption{(a) Geometry of the numerical model which includes the receiving and transmitting coils, two ferrite layers attached to each coil, a layered core-shell cylinder emulating a Li-ion cell, receiving circuit and dc-dc converters, and Ports~$1$,$2$. (b) Schematic block for the considered model having the form of a two-port circuit. The transmitter and receiver circuits include capacitors $C_{\rm T}$, $C_{\rm R}$ and the port impedances $Z_{\rm T}$, $Z_{\rm R}$, respectively.}
    \label{fig:Model}
\end{figure*}
\end{singlespace}
%______________________Figure_S1_______________________

We proceed by constructing numerical models of the transmitting and receiving coils with geometries corresponding to those widely applied in Qi-standard wireless power transfer using CST Microwave Studio. The model shown in Fig.~\ref{fig:Model}(a) includes the receiving coil (top) and the transmitting coil (bottom) supplied with ferrite layers, as well as two ports: Port~$1$ for the transmitter and Port~$2$ for the receiver, correspondingly. Inside the outline of an AA battery that is covered by the receiving coil, a layered cylinder with length $40$~mm and diameter $13.4$~mm is placed that emulates electromagnetic properties of a Li-ion cell and required electronics. Within the schematic block shown in Fig.~\ref{fig:Model}(b), single capacitors $C_{\rm T}$ and $C_{\rm R}$ are attached in series with each coil to achieve a resonant coupling at $f_{0}=100$~kHz. The port impedances $Z_{\rm T}$ and $Z_{\rm R}$ are matched, i.\,e., chosen in a way that guarantees that $S_{11}$ and $S_{22}$ are purely real and that their amplitude does not exceed $-30$~dB in the entire frequency range considered. 

The transmitting coil is implemented as a circular spiral coil having $10$ turns of a copper wire with circular cross-section, diameter of $0.4$~mm, and $0.1$~mm turn spacing. The outer diameter of the transmitting coil reaches $40$~mm. To enhance magnetic coupling, the transmitter is placed atop the ferrite substrate with diameter $42$~mm and height $1.6$~mm.

The receiving coil model represents a rectangular spiral inductor with $14$~turns of a wire consisting of two parallel copper conductors, the diameter of a single wire being $0.4$~mm. The turn spacing is $0.06$~mm, and the overall in-plane dimensions of the coil are $40 \times 30$~mm. The receiving coil is also supplied with a ferrite layer that has a thickness of $0.2$~mm. Within the numerical simulations described in the main text, the geometry of the receiving coil varies from a flat coil to cylindrical surfaces with different curvature radii $\rho$.

The receiving and the transmitting coils are surrounded by vacuum media. The wire material is copper, characterized by the electric conductivity $\sigma = 5.96 \times 10^{7}$~S/m as defined in the CST Microwave Studio materials library. For the ferrite substrates, we set the following material properties: permittivity $\varepsilon=1$ and permeability $\mu=2000$. To emulate the presence of a Li-ion cell, printed circuit boards (PCBs), and wires in the inner region surrounded by the receiving coil, we place a multilayered core-shell cylinder in this area with the following layer parameters: a ferrite layer with $D_{\rm out}=13.4$~mm and $D_{\rm in}=13.0$~mm, the first aluminum layer with $D_{\rm out}=12.8$~mm and $D_{\rm in}=10.8$~mm, a copper layer with $D_{\rm out}=10.8$~mm and $D_{\rm in}=8.8$~mm, a graphite layer with $D_{\rm out}=8.8$~mm and $D_{\rm in}=6.8$~mm, the second aluminum layer with $D_{\rm out}=6.8$~mm and $D_{\rm in}=4.8$~mm, and a copper core with $D_{\rm out}=4.8$~mm.

%______________________Figure_S2_______________________
\begin{singlespace}
\begin{figure*}[b]
    \includegraphics[width=17cm]{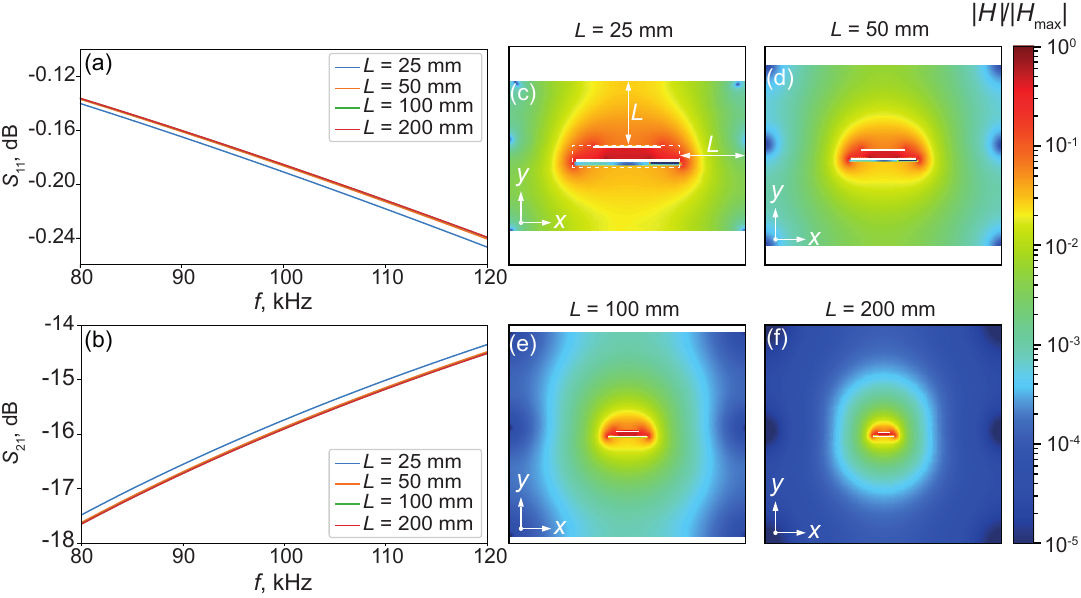}
    \caption{Comparison of numerical results for different simulation region sizes. (a) Frequency dependencies of $S_{11}$-parameters for several values of the offset $L$ between the edges of transmitting and receiving coils and the simulation region boundaries: $25$~mm, $50$~mm, $100$~mm, and $200$~mm. (b) The same as panel (a), but for $S_{21}$-parameters. (c)-(f) Magnetic field profiles in the $(xy)$-plane for the simulation box offsets (c) $25$~mm, (d) $50$~mm, (e) $100$~mm, and (f) $200$~mm, respectively. The distance between the transmitting coil and the planar receiving coil is $h=5$~mm.}
    \label{fig:Box_size_sweep}
\end{figure*}
\end{singlespace}
%______________________Figure_S2_______________________

For Qi-standard WPT operating at frequencies $100-200$~kHz, the wavelength ranges from one and a half to three kilometers, which, taken together with characteristic receiving and transmitting coils featuring sizes of several centimeters, complicates the consideration of simulation regions exceeding the wavelength. To establish the optimal simulation region size, we perform numerical calculations of $S_{11}$- and $S_{21}$-parameters and magnetic field distributions for the same configuration including the transmitting coil and the flat receiving coil separated by a $5$~mm gap, and change the offset $L$ between the model's edges and simulation region boundaries implemented as perfectly matched layers. The obtained results for $S_{11}$- and $S_{21}$-parameters demonstrate that $S$-parameters differ for $L=25$~mm and $L=50$~mm, but saturate and become hardly resolvable for offsets $L=50$~mm, $L=100$~mm, and $L=200$~mm, see Fig.~\ref{fig:Box_size_sweep}(a),(b). Moreover, magnetic fields do not considerably vanish near the simulation region boundaries for the offsets $L = 25$~mm and $L = 50$~mm [Fig.~\ref{fig:Box_size_sweep}(c),(d)], while for larger offsets $L = 100$~mm and $L = 200$~mm the magnetic field amplitudes at the boundaries are $10^{3}$ and $10^{4}$ times lower than the maximal values for the simulated models, correspondingly, Fig.~\ref{fig:Box_size_sweep}(e),(f). To find a balance between the accuracy of the simulation and the simulation time, the offset $L=100$~mm has been selected for further simulations, which corresponds to the overall simulation region sizes around $242 \times 242 \times 200$~mm$^{3}$. Note that for the described simulations, we do not add a serial capacitance to the coils. Therefore, the frequency dependencies of the $S_{11}$- and $S_{21}$-parameters do not feature any resonances in the vicinity of $f=100$~kHz.

%______________________Figure_S3_______________________
\begin{figure*}[b]
    \includegraphics[width=17cm]{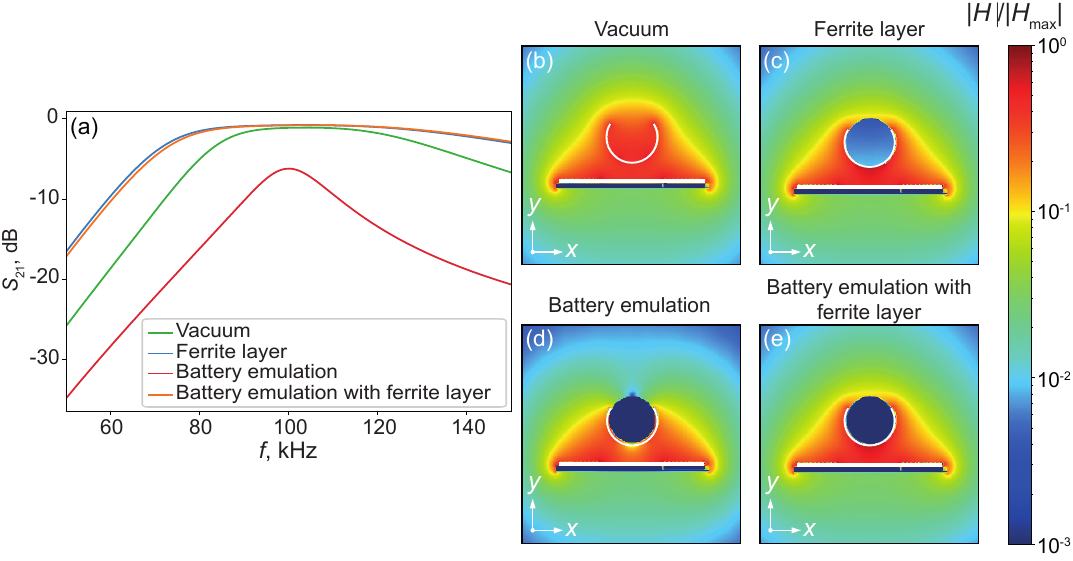}
    \caption{Comparison of numerical models with or without a ferrite layer attached to a curved receiving coil and a core-shell cylinder in the inner region of the battery outline emulating electromagnetic properties of a Li-ion cell, printed circuit boards, and wires. (a) Frequency dependencies of $S_{21}$-parameters for the four considered configurations. (b)-(e) Magnetic field profiles in the $(xy)$-plane for the following cases: (b) without both a ferrite layer and components emulation, (c) with a ferrite layer, but without components emulation, (d) without a ferrite layer, but with components emulation, and (e) with both a ferrite layer and components emulation. In all cases, the models with $\rho=7$~mm and $h=5$~mm are considered.}
    \label{fig:Ferrite_and_battery}
\end{figure*}
%______________________Figure_S3_______________________

Next, we address the effects of components such as a Li-ion cell, PCBs, and wires placed near the receiving coil on the magnetic field distributions and $S_{21}$-parameters. In particular, we consider a cylindrically bent receiving coil with curvature radius $\rho = 7$~mm, the same coil supplied with a thin ferrite layer, the coil with a layered cylinder emulating other components of the battery, and, finally, the receiving coil with both a cylinder and a ferrite layer. The obtained $S_{21}$-parameters demonstrate that the presence of the cylinder deteriorates the transmission coefficient between the transmitting and the receiving coils, Fig.~\ref{fig:Ferrite_and_battery}(a). Nevertheless, incorporating a thin ferrite layer between the coil and the cylinder provides sufficient magnetic shielding and considerably improves the coupling efficiency. As seen from the magnetic field profiles in Fig.~\ref{fig:Ferrite_and_battery}(b)-(e), when the ferrite layer is introduced, the magnetic field profiles demonstrate a more pronounced concentration between the centers of the receiving and the transmitting coils, thus increasing their coupling efficiency, Fig.~\ref{fig:Ferrite_and_battery}(b),(c). In turn, the addition of battery components to the receiving coil in the absence of the ferrite layer leads to a local decrease in the magnetic field amplitude, as shown in Fig.~\ref{fig:Ferrite_and_battery}(d). Finally, the results for the model that combines the presence of components and the ferrite layer [Fig.~\ref{fig:Ferrite_and_battery}(e)] closely resemble those obtained for the receiving coil with a ferrite layer and without a cylinder, highlighting the efficiency of magnetic shielding.

%___________________S3_Receive_Coils____________________
\section*{Supplementary Note 3. Comparison of receiving coils}

%______________________Figure_S4_______________________
\begin{figure*}[b]
    \includegraphics[width=12cm]{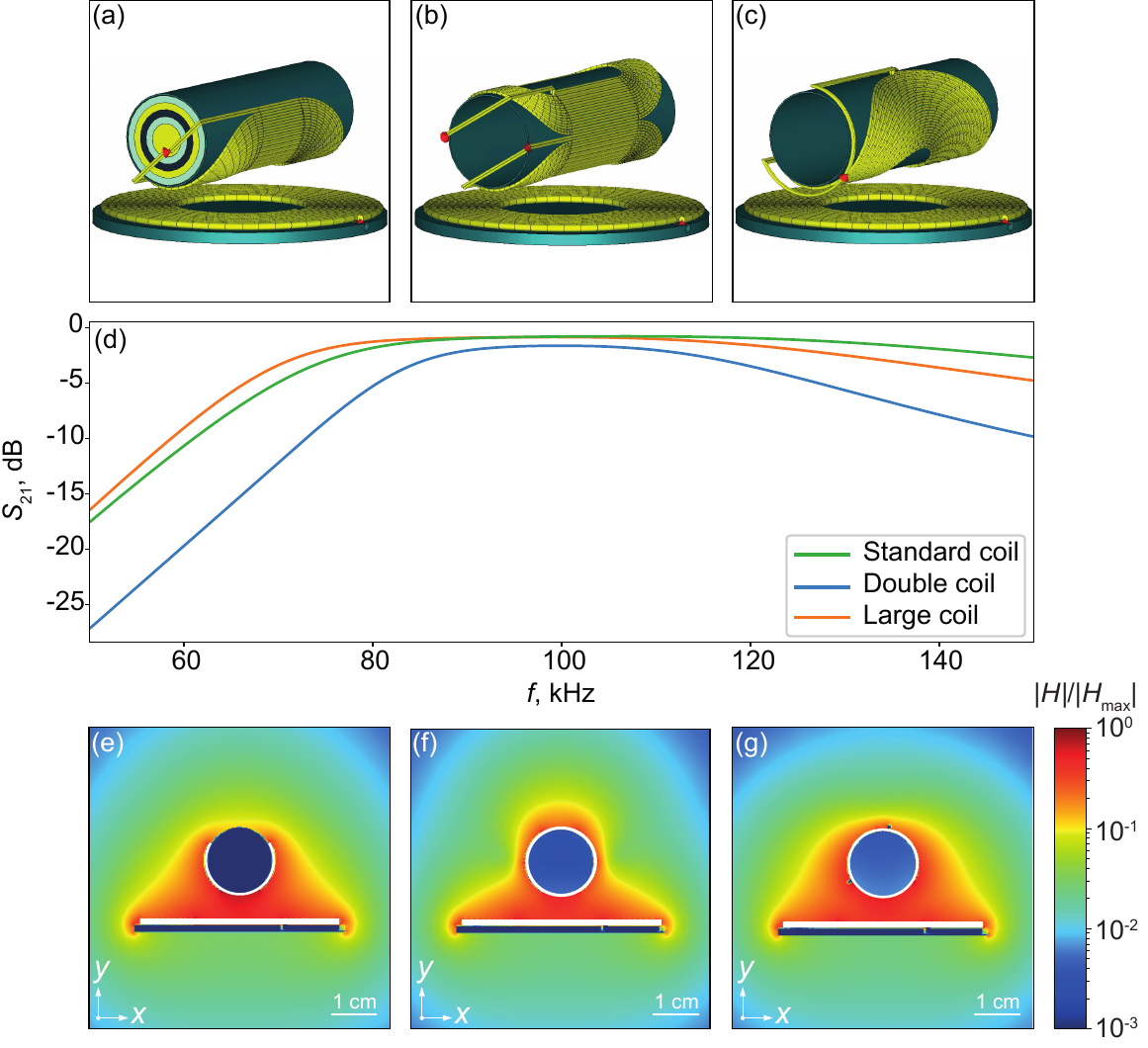}
    \caption{Comparison of three considered receiving coil designs: (a) a single coil having geometric parameters characteristic of commercial Qi coils and covering half of an AA battery surface; (b) two such coils covering full battery surface; and (c) an enlarged individual coil covering full battery surface. In all three cases, the coils are supplied with a ferrite layer. (d) $S_{21}$-parameters for the three considered coil types located above the transmitting coil at the height of $5$~mm. (e)-(g) The distributions of the magnetic field in $(xy)$ plane for the center of the battery $z=0$ corresponding to (e) standard, (f) double, and (g) enlarged coils.}
    \label{fig:Coils}
\end{figure*}
%______________________Figure_S4_______________________

In this Note, we compare several receiving coil designs to establish the most efficient ones that can serve as starting points in further engineering optimization. We examine the following designs: the curved standard Qi receiving coil considered in the main text that covers more than half of the battery surface [Fig.~\ref{fig:Coils}(a)], a combined coil consisting of two coils that together cover the whole battery surface [Fig.~\ref{fig:Coils}(b)], and an enlarged individual coil that covers the whole battery surface [Fig.~\ref{fig:Coils}(c)]. As seen from frequency dependencies of the $S_{21}$-parameters in Fig.~\ref{fig:Coils}(d), the double coil renders less efficient compared to the standard and enlarged coils, which demonstrate nearly the same coupling with the transmitter, with the enlarged coil being slightly more efficient at low frequencies and the standard coil outperforming the rest in the high frequency region. A comparison of magnetic field distributions in Fig.~\ref{fig:Coils}(e)-(g) demonstrates that the double coil in Fig.~\ref{fig:Coils}(f) features two areas of pronounced magnetic field localization at the centers of two coils, in contrast to the standard coil in Fig.~\ref{fig:Coils}(e), while the field distribution for the enlarged coil in Fig.~\ref{fig:Coils}(g) resembles the one for the standard coil. Thus, counterintuitively, considering designs with several coils or maximizing the battery area covered by the receiving coil may not lead to the increase in WPT efficiency, which, as shown in Fig.~5 of the main text, remains nearly unchanged upon bending the receiving coil compared to the standard planar receiver.

%____________________S4_Experiments____________________
\section*{Supplementary Note 4. Measurement of $S$-parameters}

%______________________Figure_S5_______________________
\begin{figure*}[b]
    \centering
    \includegraphics[width=17cm]{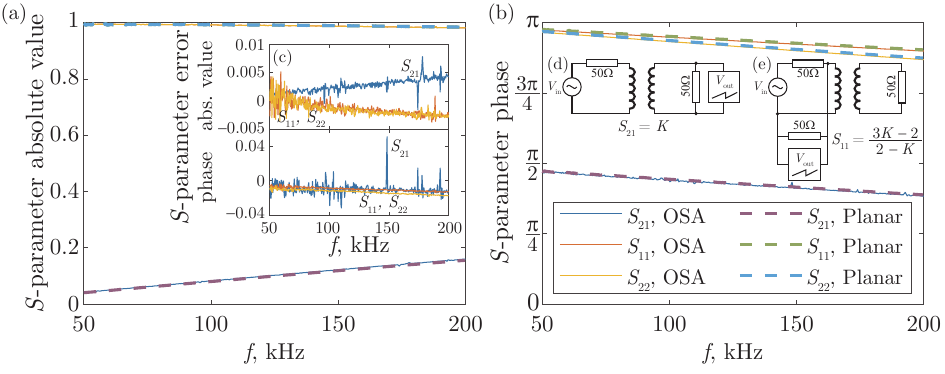}
    \caption{Amplitudes (a) and phases (b) of the reflection coefficients $S_{11}$, $S_{22}$ and transmission coefficient $S_{21}$ measured experimentally with OSA103 Mini (solid lines) and with Planar S5048 (dashed lines). The transmission coefficient $S_{12}$ is equal to $S_{21}$ by the reciprocity theorem and not shown. Inset (c) shows the differences between the $S$ parameters measured by the aforementioned devices. Inset (d) shows the circuit used to measure $S_{12}$ and $S_{21}$ parameters with the OSA103 Mini device. Inset (e) shows the circuit used to measure $S_{11}$ and $S_{22}$ coefficients with the OSA103 Mini device.}
    \label{fig:OSA_verification}
\end{figure*}
%______________________Figure_S5_______________________

To obtain the $S$-parameters in the kHz range, we utilize OSA103 Mini open-source platform for electrical measurements featuring acquisition of amplitude and phase frequency responses in the $100$~Hz to $100$~MHz range. As this device can only work as a scalar network analyzer capable of measuring the transmission coefficient $S_{21}$ of a four-pole circuit, we use the following approach to obtain the $S$-parameters by indirect measurement. To obtain the transmission coefficient $S_{21}$, we use a setup shown in Fig.~\ref{fig:OSA_verification}(d). The $50$~Ohm resistors in both the input and output ports are built into the OSA103 Mini device. Before measurements, a calibration was performed by replacing the coupled coils with a direct connection of the input and output ports of the device by a $10$-centimeter RJ142 coaxial cable. After applying the calibration, the OSA103 Mini device becomes configured in such a way that the complex transmission coefficient $K$ obtained from the device for the direct output-to-input port connection becomes unity. By connecting the transmitting coil to the output port and the receiving coil to the input port as shown in Fig.~\ref{fig:OSA_verification}(d), one obtains the complex transmission coefficient $K$, which is equal to the transmission coefficient $S_{21}$. To obtain the reverse transmission coefficient $S_{12}$, the receiving coil is connected to the output port, and the transmitting coil is connected to the input port. We note that all the circuits for which we obtain the $S$-parameters are reciprocal, thus $S_{12}$ is always equal to $S_{21}$. In our studies, we also performed measurements to obtain $S_{12}$. However, since it is equal to $S_{21}$ in each case, we do not show it for clarity.

To obtain the reflection coefficients $S_{11}$ and $S_{22}$ without the use of directional couplers, which limit the accessible frequency range from the low-frequency side, we perform an indirect measurements using the circuit shown in Fig.~\ref{fig:OSA_verification}(e). A tee adapter is installed onto the output port of the OSA103 Mini device, with one of its outputs connected to the input port using an RJ142 coaxial cable and the other one to the input port of the four-pole studied, which is the transmitting coil in the case shown in Fig.~\ref{fig:OSA_verification}(e). The output port of the four-pole under test is terminated by a 50~Ohm dummy load. By writing down the Kirchhoff's rules for such a circuit, one can obtain the equation connecting the complex transmission coefficient $K$ obtained from the OSA103 Mini device to the reflection coefficient $S_{11}$ of the four-pole:
\begin{equation}
    S_{11} = \frac{3K - 2}{2 - K}.
\end{equation}
To obtain the $S_{22}$ reflection coefficient, the receiving coil is connected to the output port, while the transmitting coil is terminated with the dummy load.

To verify our indirect measurements, we compare the $S$-parameters obtained using our approach with the $S$-parameters obtained from the Planar S5048 vector network analyzer, which has a 20~kHz to 4.8~GHz range. The four-pole used for this verification consisted of coupled planar transmitting and receiving coils placed 6~mm away from each other. The $S_{11}$, $S_{22}$, and $S_{21}$ parameters obtained with both devices in the 50~kHz to 200~kHz range for two planar coupled coils are shown in Fig.~\ref{fig:OSA_verification}(a),(b), while the inset in Fig.~\ref{fig:OSA_verification}(c) shows the difference between the values obtained by both devices. % The relative error in the given range does not exceed 10\%.

%_________________________S5_Coupling_setup____________________________
\section*{Supplementary Note 5. Experimental setup for studies of the coupling between the receiver and transmitters}

\begin{figure*}[tb]
    \centering
    \includegraphics[width=8.5cm]{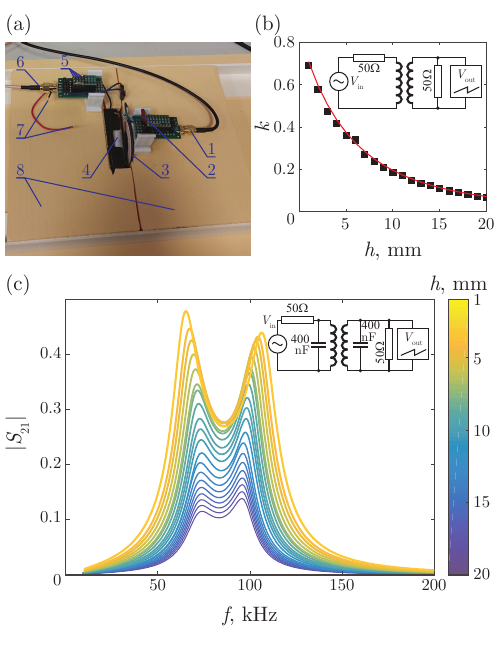}
    \caption{(a) Experimental setup for $S$-parameters measurement, consisting of (1) the VNA output port, (2) a 400~nF polymer film capacitor which can be assembled into a serial or parallel $LC$-circuit with the transmitting coil or bypassed using jumpers, (3) the transmitting coil, (4) the receiving coil inside an AA battery-sized photopolymer resin casing mounted inside an AA battery holder, (5) four ceramic capacitors $68$~nF each connected in parallel which can be assembled into a serial or parallel $LC$-circuit with the receiving coil or bypassed using jumpers, and (6) the VNA input port. Battery holder output leads (7) are attached to provide the battery output voltage measurement. The setup is mounted on a radio-transparent extruded polystyrene foam foundation (8) which can be moved in one direction to change the distance $h$ between the receiving and transmitting coils. (b) The coil-to-coil coupling $k$ dependence on the distance $h$ between the transmitting and receiving coils. The circuit diagram is shown in the inset. (c) Experimentally measured dependence of $S_{21}$-parameters on the distance between the transmitting and receiving coils varied in the range from $1$ to $10$~mm. The inset shows the circuit diagram.}
    \label{fig:Sparameters}
\end{figure*}

To perform the $S$-parameter measurement, the setup shown in Fig.~\ref{fig:Sparameters}(a) was used. It features two SMA connectors (1) and (6) used to connect the setup to the output and input ports of a vector network analyzer (VNA), PCBs (2) and (5) with tank circuit capacitors and jumper terminals, the transmitting coil (3), and the receiving coil (4). The transmitter and receiver parts of the setup are mounted on radio-transparent extruded polystyrene foam (XPS) foundations (8) that can be moved in one direction to vary the distance $h$ between the receiving and transmitting coils.

The transmitting coil (3) is a spiral planar coil consisting of $10$ turns of PTFE-insulated stranded copper wire with $1$~mm diameter (the outer diameter of the coil is $42$~mm, the inner diameter is $22$~mm) placed on a disk-shaped ferrite layer with $45.5$~mm diameter and $0.8$~mm thickness. The capacitor (2) used with the transmitting coil is a single \mbox{CBB21}-\mbox{404J}-\mbox{100V} polypropylene film capacitor with a nominal capacitance of $400~\text{nF}\pm 5\%$. Both the transmitting coil (3) and the capacitor (2) have been extracted from a disassembled Qi charger.

The receiving coil is a planar coil in the shape of a rounded rectangle with length $40.5$~mm and width $30.5$~mm, consisting of $14$ turns of a wire made of two parallel monolithic enameled copper conductors with circular cross-section and diameter of $0.4$~mm, which are soldered together at their ends, with an inner $22.5 \times 10.5~\text{mm}^2$ empty space. The coil is bent around a cylindrical surface with the axis parallel to the large side of the coil and the curvature radius $\rho=7$~mm. The coil is then enclosed in an AA battery-sized casing manufactured using Anycubic Photon SLA 3D printer and backed with a $41 \times 28~\text{mm}^2$ flexible ferrite sheet with $0.1$~mm thickness from the interior side. As seen from the numerical simulations, the presence of components inside the battery does not considerably change the field distributions and $S$-parameters due to the magnetic field shielding by a ferrite layer. To confirm this, a Robiton 10440-sized $350$~mAh Li-ion cell was installed inside the battery casing to simulate the presence of the aforementioned components. The casing containing the coil was installed in an AA battery holder mounted on a movable XPS foundation. The receiver tank capacitor assembly (5) consists of four Murata \mbox{GRM31C5C1H104JA01L} C0G ceramic capacitors, each having $100~\text{nF}\pm 5\%$ capacitance, connected in parallel, adding up to a total capacitance of $400$~nF.

The capacitors beside both the receiving and the transmitting coils were mounted on PCBs that also housed the SMA connectors and the jumper terminals, which allow one to assemble a parallel $LC$ circuit [inset of Fig.~\ref{fig:Sparameters}(c)], a serial $LC$ circuit, or to connect the coil directly to the port [inset of Fig.~\ref{fig:Sparameters}(b)], depending on the jumper placement.

%______________________Figure_S6_______________________
%______________________Figure_S6_______________________

%______________________Figure_S7_______________________
\begin{figure*}[p]
    \centering
    \includegraphics[width=15cm]{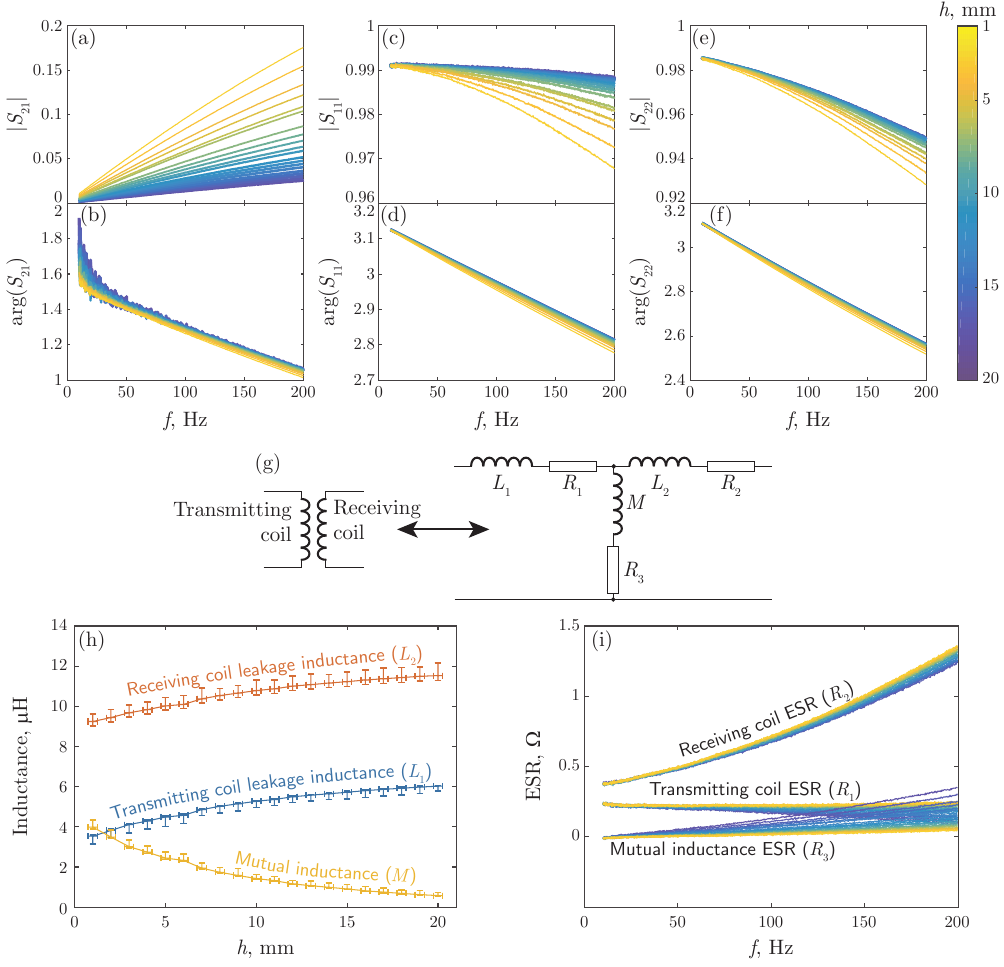}
    \caption{Experimentally measured $S$-parameters of coupled transmitting and receiving coils for the distance $h$ between the transmitting and the receiving coils varied from 1~mm to 20~mm: (a) the amplitude of the transmission coefficient $S_{21}$ and (b) its phase, (c) the amplitude of the reflection coefficient from the transmitting coil $S_{11}$ and (d) its phase, (e) the amplitude of the reflection coefficient from the receiving coil $S_{22}$ and (f) its phase. The transmission coefficient $S_{12}$ is equal to $S_{21}$ by the reciprocity theorem and not shown. (g) Equivalent circuit diagram of the coupled transmitting and receiving coils. (h) The dependencies of the mutual ($M$) and leakage ($L_{1,2}$) inductances of the transmitting and receiving coils on the distance $h$ between the coils. The vertical error bars show the variation of the inductance in the 1~kHz to 200~kHz frequency range. The lines connect the average inductance values over the aforementioned frequency range. The horizontal error bars indicate the positioning error. (i) Equivalent serial resistance (ESR) of the equivalent inductances corresponding to the transmitting and receiving coils leakage and mutual inductances, shown as a frequency dependence for the distance $h$ between the coils varied from 1~mm to 20~mm.}
    \label{fig:CoilSparameters}
\end{figure*}
%______________________Figure_S7_______________________

%_____________________S6_Experiments_coupling________________________
\section*{Supplementary Note 6. Experimental studies of coupling between the receiving and transmitting resonant coils}

We proceed with the experimental studies of the effective inductances of coupled coils. We start by obtaining the full set of $S$-parameters for coupled transmitting and receiving coils placed at distance $h$ away from each other, where the transmitting coil is a planar spiral coil consisting of 10 turns of $1$~mm-thick copper wire (with outer diameter $42$~mm and inner diameter $22$~mm) placed on a disk-shaped ferrite foundation (with diameter $45.5$~mm and thickness $0.8$~mm), and the receiving coil is a $14$-turn bent $40 \times 30~\text{mm}^2$ coil with the curvature radius $\rho=7$~mm, enclosed in an AA battery-sized photopolymer resin case and backed with a $41 \times 28~\text{mm}^2$ flex ferrite sheet with $0.1$~mm thickness from the interior side. The resulting $S$-parameters for $h$ varied from $1$~mm to $20$~mm are shown in Fig.~\ref{fig:CoilSparameters}(a)-(f).

Imperfectly-coupled inductances can be described with the equivalent circuit shown in Fig.~\ref{fig:CoilSparameters}(g)~\cite{1976_Daniels}. We use the obtained $S$ parameters to extract leakage inductances $L_1$ and $L_2$ and their equivalent series resistances (ESR) $R_1$ and $R_2$, as well as mutual inductance $M$ and its ESR $R_3$. We start by converting the $S$-parameter matrix to the $ABCD$ matrix using the formulae~\cite{1994_Frickey}
\begin{equation}
\label{eq:StoABCD}
\begin{aligned}
    A &= \frac{(Z_{01}^{*} + S_{11}Z_{01})(1 - S_{22}) + S_{12}S_{21}Z_{01}}{2S_{21} \sqrt{R_{01}R_{02}}}, &
    B &= \frac{(Z_{01}^{*} + S_{11}Z_{01})(Z_{02}^{*} + S_{22}Z_{02}) - S_{12}S_{21}Z_{01}Z_{02}}{2S_{21} \sqrt{R_{01}R_{02}}}, \\
    C &= \frac{(1 - S_{11})(1 - S_{22}) - S_{12}S_{21}}{2S_{21} \sqrt{R_{01}R_{02}}}, &
    D &= \frac{(1 - S_{11})(Z_{02}^{*} + S_{22}Z_{02}) + S_{12}S_{21}Z_{02}}{2S_{21} \sqrt{R_{01}R_{02}}},
\end{aligned}
\end{equation}
where $Z_{01} = 50~\mathrm{\Omega}$ is the source output impedance, $Z_{02} = 50~\mathrm{\Omega}$ is the load impedance, $R_{01} = \mathrm{Re}\,Z_{01}$ and $R_{02} = \mathrm{Re}\,Z_{02}$. Next, we use the $ABCD$ matrix for a four-pole T circuit~\cite{1987_Fusco}:
\begin{equation}
\label{eq:ABCDtoSteinmetz}
    \begin{aligned}
        A &= 1 + \frac{Z_1}{Z_3}, &
        B &= Z_1 + Z_2 + \frac{Z_1Z_2}{Z_3}, \\
        C &= \frac{1}{Z_3}, &
        D &= 1 + \frac{Z_2}{Z_3},
    \end{aligned}
\end{equation}
where
\begin{equation}
\label{eq:SteinmetzImpedances}
    \begin{aligned}
        Z_1 &= i\omega L_1 + R_1, &
        Z_2 &= i\omega L_2 + R_2, &
        Z_3 &= i\omega M + R_3.
    \end{aligned}
\end{equation}
Combining Eqs.~\eqref{eq:StoABCD} to \eqref{eq:SteinmetzImpedances}, we obtain a closed system that expresses the leakage and mutual inductances and the corresponding ESR through the experimentally-measured $S$ parameters. Solving these equations for each frequency yields the values shown in Fig.~\ref{fig:CoilSparameters}(h),(i). Note that the inductances do not demonstrate any significant change with frequency. Therefore, we show them in Fig.~\ref{fig:CoilSparameters}(h) as dots with error bars that correspond to their variation in the whole 1~kHz to 200~kHz range. However, these inductances are significantly dependent on distance $h$. We also note that the self-inductances are almost constant for all distances. The self-inductance of the transmitting coil $L_{\text{Tx}} = L_1 + aM \simeq 7~\mu\text{H}$, where $a = 10/14$ is the winding turns ratio, while the self-inductance of the receiving coil $L_{\text{Rx}} = L_2 + M/a \simeq 12~\mu\text{H}$. In contrast, the resistances shown in Fig.~\ref{fig:CoilSparameters}(i) considerably depend on the frequency, while their dependence on the distance between the coils is weak. The apparent increase of the mutual inductance ESR $R_3$ at larger distances may be related to absorption of the stray ac magnetic field by surrounding objects, as well as the magnetic field screening by the ferrite sheet becoming less efficient.

The coupling $k = M/\sqrt{L_1 L_2}$ demonstrates a hyperbolic dependency $k = [(h + h_0)/h_1]^{-\alpha}$ on the distance $h$, which is shown in Fig.~\ref{fig:Sparameters}(b). A least squares fit yields the following parameter values: $h_0 = 16.5~\text{mm}$, $h_1 = 15.7~\text{mm}$, $\alpha = 3.1$.

%______________________Figure_S8_______________________
\begin{figure*}[tb]
    \centering
    \includegraphics[width=17cm]{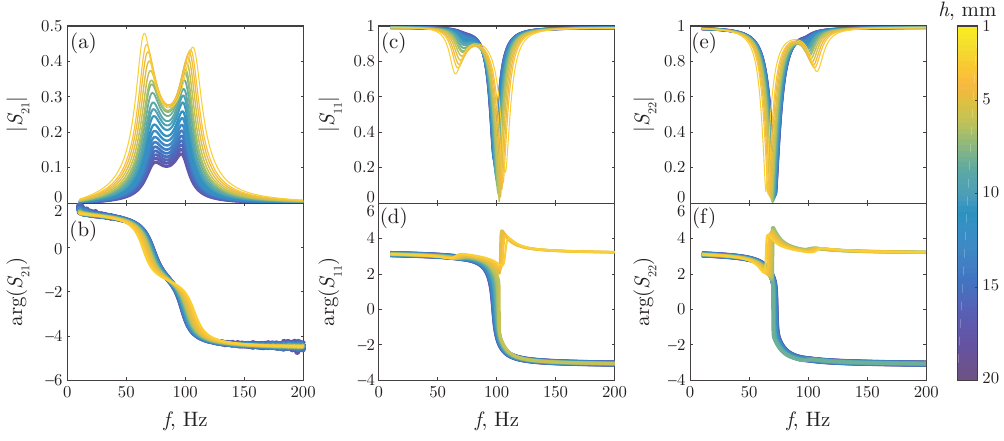}
    \caption{Experimentally measured $S$-parameters of coupled parallel contours for the distance $h$ between the transmitting and the receiving coils varied from $1$~mm to $20$~mm: (a) amplitude of the transmission coefficient $S_{21}$ and (b) its phase, (c) amplitude of the reflection coefficient from the transmitting coil $S_{11}$ and (d) its phase, (e) amplitude of the reflection coefficient from the receiving coil $S_{22}$ and (f) its phase. The transmission coefficient $S_{12}$ is equal to $S_{21}$ by the reciprocity theorem and not shown.}
    \label{fig:SparametersFull}
\end{figure*}
%______________________Figure_S8_______________________

Finally, we add capacitors to the transmitting and receiving coils and study the $S$-parameters of the resulting coupled parallel $LC$ circuits. Figure~\ref{fig:Sparameters}(c) shows the dependence of the transmission coefficient $S_{21}$ from the transmitting coil to the receiving coil for the distance $h$ varied from $1$~mm to $20$~mm, while Figure~\ref{fig:SparametersFull} shows the complete set of $S$-parameters. First, it is seen that there are two clearly resolved resonant peaks in the dependence of $|S_{21}|(f)$ on the frequency $f$. Their frequencies change considerably with distance $h$ between the coils at low distances, but eventually saturate at greater spacings; see Fig.~\ref{fig:Sparameters}(c). The appearance of a coupling-dependent resonant peak splitting indicates a hybridization of individual coils' resonant states, characteristic of strong-coupling regime. Such systems typically display a fork-like dependence of the resonant frequencies on the coupling strength, with the transition to the weak-coupling regime marked by merging of two resonances into one. However, in our case, the $LC$ circuits have different resonant frequencies ($100$~kHz for the transmitting coil and $70$~kHz for the receiving coil). Thus, rather than a splitting, the distance dependency of the resonant frequencies demonstrates an anticrossing, which is yet another manifestation of the strong coupling regime. Note that only the decreasing region of the high-frequency resonant peak is used for power transmission to monotonously tune the transmission rate with frequency according to the Qi standard. The frequency of the higher resonance peak remains in the range defined by the Qi standard for all considered distances between the coils.

%________________________S7_Magnetic_field____________________________
\section*{Supplementary Note 7. Evaluation of the WPT magnetic field influence on internal components}

\begin{figure*}[b]
    \centering
    \includegraphics[]{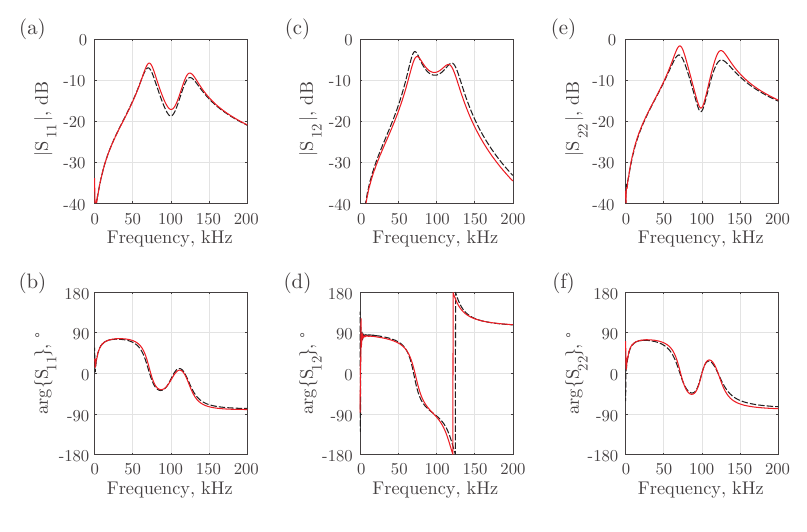}
    \caption{Experimentally measured $S$-parameters of coupled parallel contours for the distance $h=1~\text{mm}$ between the transmitting and receiving coils for an empty battery with the receiving coil backed with ferrite (dashed curves) and for a battery with all its internal components (solid curves): (a) amplitude of the reflection coefficient from the transmitting coil $S_{11}$ and (b) its phase, (c) amplitude of the transmission coefficient $S_{21}$ and (d) its phase, (e) amplitude of the reflection coefficient from the receiving coil $S_{22}$ and (f) its phase. The transmission coefficient $S_{21}$ is equal to $S_{12}$ by the reciprocity theorem and not shown.}
    \label{fig:SparametersFerrite}
\end{figure*}

To experimentally evaluate the influence of the WPT magnetic field on the internal components of the battery and vise versa, we compare the scattering parameters of a battery without all of its internal components but the receiving coil, ferrite shielding, and tank capacitor, and a battery with all of its internal components. The scattering parameters obtained for the separation $h = 1~\text{mm}$ between the transmitting and receiving coils are shown in Fig.~\ref{fig:SparametersFerrite}.

%______________________Figure_S9_______________________
%______________________Figure_S9_______________________

Due to the ferrite backing of the receiving coil, the mutual influence between the WPT magnetic field and the internal components of the battery is mostly alleviated, which is confirmed by the agreement between the spectra in Fig.~\ref{fig:SparametersFerrite}. As expected, the reflection coefficient $S_{22}$ from the battery in Fig.~\ref{fig:SparametersFerrite}(e),(f) experiences the largest change with the introduction of the internal components reaching 3~dB. The amplitude of the transmission coefficient peak in the working frequency range ($f > 100~\text{kHz}$) is -5.9~dB at 120.4~kHz for the empty battery and -6.2~dB at 106.2~kHz for the battery with internal components. This change is small enough to consider the influence of the internal components negligible.

%______________________S8_Rectifier_topology______________________
\section*{Supplementary Note 8. Choice of the rectifier topology}
\label{sec:Rectifiers}

%______________________Figure_S10_______________________
\begin{figure*}[tb]
    \centering
    \includegraphics[]{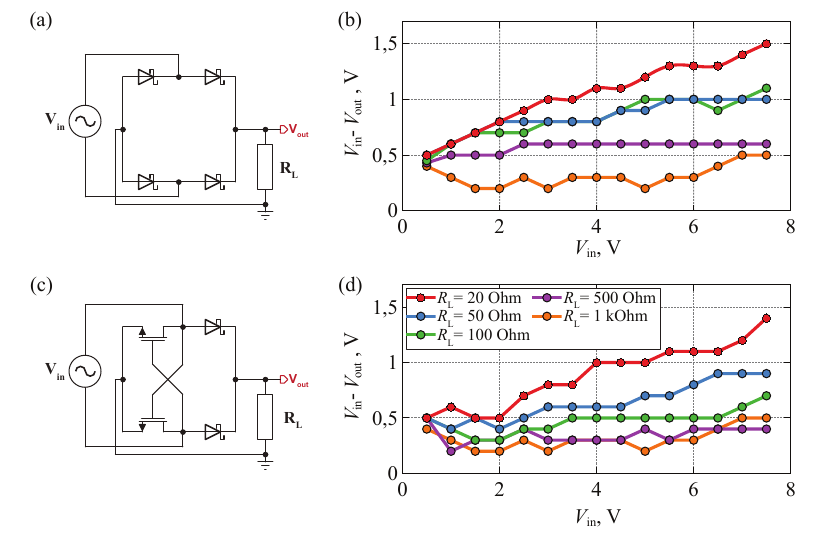}
    \caption{Experimentally measured rectifier voltage drops. (a) Schematic diagram of a full-bridge rectifier consisting of four SK14 Schottky diodes. (b) Voltage drops of the aforementioned rectifier measured at $f = 100~\text{kHz}$ as the dependence of the amplitude $V_{\text{in}}$ of the input AC voltage for load resistances $R_L$ from 20~Ohm to 1~kOhm. (c) Schematic diagram of a full-bridge rectifier consisting of two SK14 Schottky diodes and two Si2306 n-MOSFETs. (d) Voltage drops of the aforementioned rectifier at the same conditions.}
    \label{fig:Rectifiers}
\end{figure*}
%______________________Figure_S10_______________________

To ensure better efficiency, a rectifier with a voltage drop as low as possible should be selected. A zero-voltage switching rectifier would be ideal, but it requires active switching circuitry and an auxiliary power supply, contradicting the requirement to keep the receiver circuitry compact. Two rectifier topologies offer a compromise between circuit simplicity and low voltage drop: a full-bridge rectifier with Schottky diodes shown in Fig.~\ref{fig:Rectifiers}(a) and a full-bridge rectifier where two of the diodes are replaced with n-MOSFETs, as shown in Fig.~\ref{fig:Rectifiers}(c). In the latter rectifier, n-MOSFETs initially conduct current through their body diodes, providing self-biasing. As a result, the gate voltage increases and opens the respective MOSFET, providing a lower voltage drop.

To compare the efficiency of these two topologies, measurements of their respective voltage drops have been performed using an Owon DGE2070 DDS generator connected to an AC voltage follower circuit made from an AD823 operational amplifier driving two complementary bipolar Darlington transistors (KT8130A and KT8131A, analogs of BD647 and BD875) to feed AC currents of up to 1~A into the tested rectifier. Voltage drop measurements have been performed for different loads $R_{\text{L}}$ connected to the rectifier output. The peak values of the AC signal at the input $V_{\text{in}}$ and the output $V_{\text{out}}$ of a rectifier were measured using an Owon DS1102 oscilloscope.

Figure~\ref{fig:Rectifiers}(b) shows the voltage drop ($V_{\text{in}} - V_{\text{out}}$) on the full-bridge rectifier consisting of four SK14 Schottky diodes as a function of the input signal amplitude $V_{\text{in}}$ at $f = 100$~kHz. At low load currents, the forward voltage drop remains almost independent of the input amplitude and is approximately equal to two times the forward voltage of the Schottky diode. However, the series resistance of the diodes comes into play at high load currents, leading to forward voltage drops of up to 1.5~V for $R_{\text{L}} = 20~\text{Ohm}$. Figure~\ref{fig:Rectifiers}(d) shows the voltage drop on the full-bridge rectifier with two diodes replaced by Si2306 n-MOSFETs. At low load currents, its performance is the same as that of a Schottky diode rectifier, while it offers better performance at large load currents due to low on-state resistance of n-MOSFETs ($R_{\text{ds}}^{\text{(on)}} = 38~\text{mOhm}$), leading to smaller voltage drops. Therefore, the latter rectifier appears to be more suitable for the Qi receiver.

%_______________________S9_Compatibility___________________________
\section*{Supplementary Note 9. Compatibility testing with different charging stations}

%______________________Figure_S11_______________________
\begin{figure*}[b]
    \centering
    \includegraphics[]{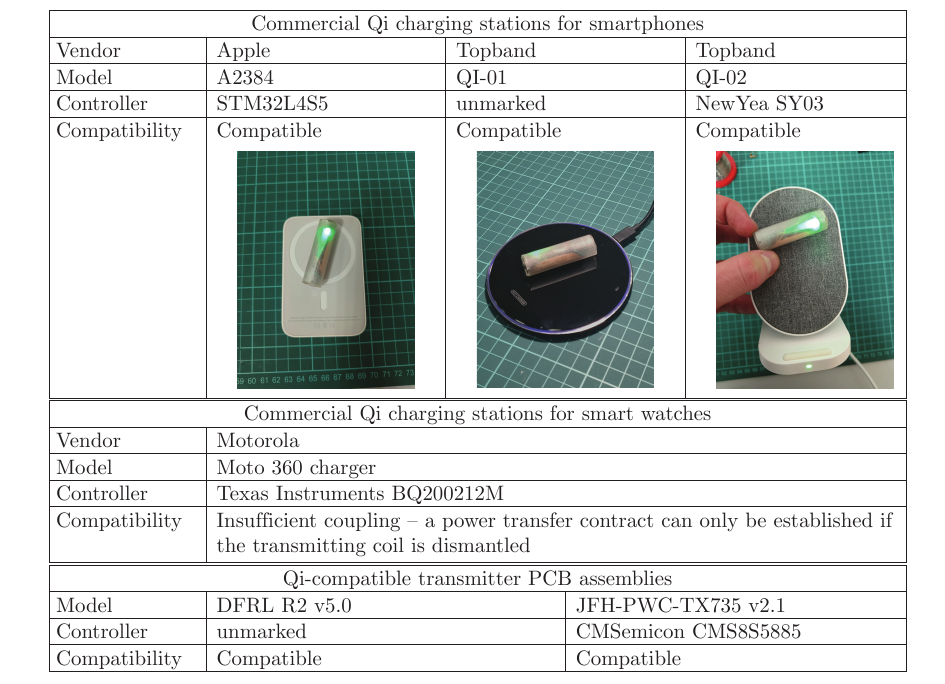}
    \caption{Compatibility with commercial Qi charger stations from various vendors as well as Qi-compatible transmitter printed circuit board (PCB) assemblies. An established power transfer contract between commercial charger stations and the AA battery prototype is indicated in the photographs provided in the insets by the green LED glowing in the prototype and green or blue LEDs glowing in the chargers.}
    \label{fig:ChargerCompatibility}
\end{figure*}
%______________________Figure_S11_______________________

To ensure compatibility with the Qi standard, we tested our prototype with a number of commercial Qi chargers: three commercial Qi-compliant chargers for smartphones, one commercial Qi charger for a smart watch, and two Qi-compatible transmitter modules that are sold uncased. The vendors, models, and controller chips (where identifiable) of the chargers are provided in the table shown in Fig.~\ref{fig:ChargerCompatibility}, as well as the results of compatibility testing. Our prototype and the charging station are considered compatible if they establish a power transfer contract, which is indicated by light-emitting diodes in both our prototype and the charging station. With the only exception of the Motorola Moto 360 smartwatch charger, which, due to its transmitting coil placed deep in the casing, only established a power transfer contract with our prototype when it was dismantled and its transmitting coil was put against the AA battery prototype, all of the tested charging stations were found to be compatible.

%______________________S10_Thermal_studies________________________
\section*{Supplementary Note 10. Study of prototype heating during charging}

%______________________Figure_S12_______________________
\begin{figure*}[tb]
    \centering
    \includegraphics[]{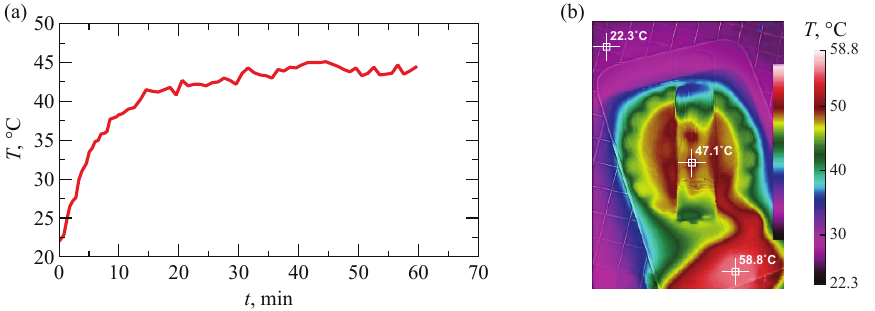}
    \caption{Experimentally measured heating of the AA battery prototype during wireless charging. (a) Time dependence of the maximum temperature $T$ observed inside the battery. (b) Heat map of the battery prototype showing the point of its maximal temperature.}
    \label{fig:Temperature}
\end{figure*}
%______________________Figure_S12_______________________

Thermal analysis during wireless charging of the AA battery prototype is crucial given the compact design and potential heat buildup. Therefore, we measured the temperature distribution within the prototype when charging from an Apple A2384 MagSafe charging station with a UNI-T UTi160S thermographic camera. Figure~\ref{fig:Temperature} shows the time dependence of the highest temperature observed in our prototype and an example of the temperature map. The room temperature at the time of measurement was $22^{\circ}~\text{C}$. The prototype temperature was found to be the highest near the Schottky diodes and the DC-DC converter, reaching as high as $47^{\circ}~\text{C}$. However, after an initial period of growth, the temperature remains stable throughout the entire charging time.

%__________________________S11_Efficiency____________________________
\section*{Supplementary Note 11. Evaluation of the system efficiency}

%______________________Figure_S13_______________________
\begin{figure*}[p]
    \centering
    \includegraphics[]{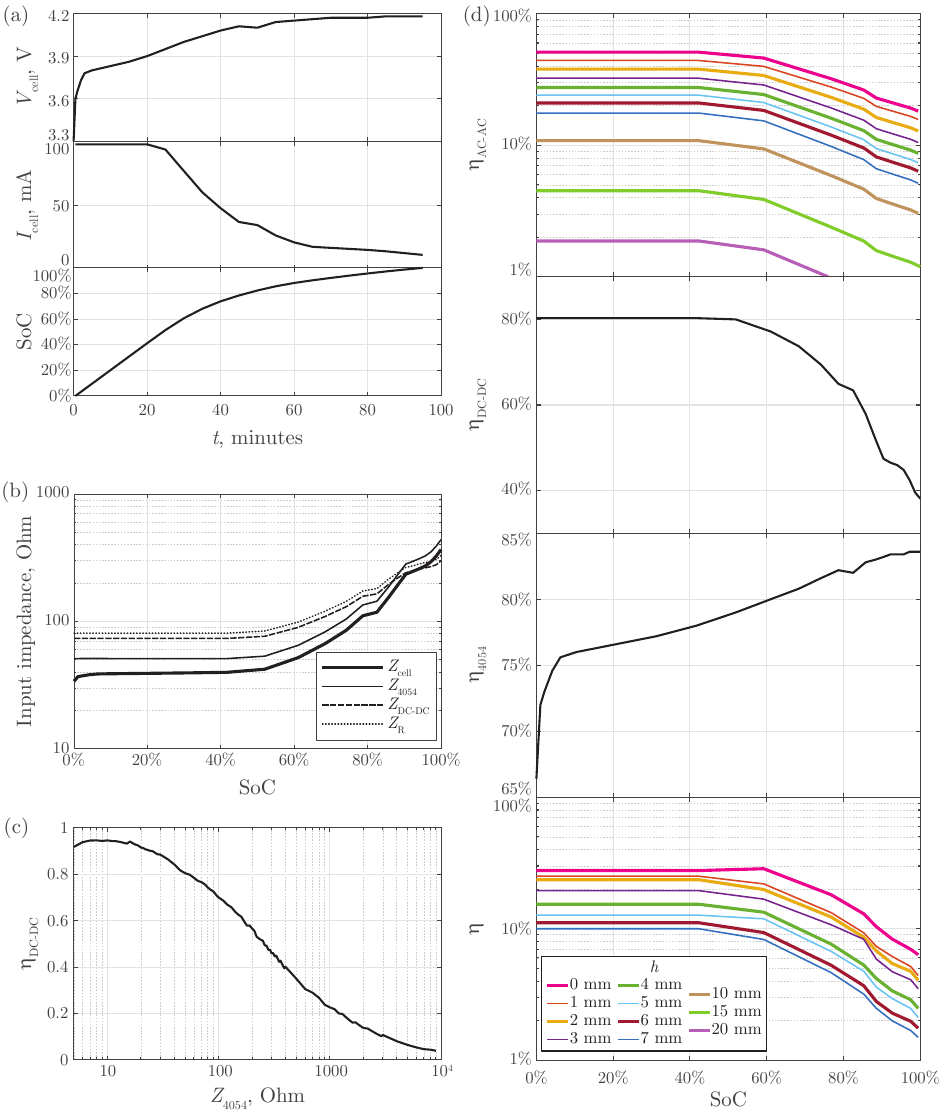}
    \caption{Efficiency of battery wireless charging and its various parts: (a) Time dependencies of the Li-ion cell voltage $V_{\text{cell}}$, charge current $I_{\text{cell}}$, and the corresponding state of charge (SoC). (b) Effective input impedances of the Li-ion cell ($Z_{\text{cell}}$), the STC4054 battery charger IC ($Z_{4054}$), the 5~V DC-DC converter ($Z_{\text{DC-DC}}$), and the rectifier ($Z_{\text{R}}$) for different SoC. (c) Efficiency of the 5~V DC-DC converter ($\eta_{\text{DC-DC}}$) as the dependence of its load $Z_{4054}$. (d) Efficiency dependence on the SoC:
    AC-AC efficiency $\eta_{\text{AC-AC}}$ of the wireless power transfer for different distances $h$ between the transmitting and the receiving coil, the DC-DC efficiency $\eta_{\text{DC-DC}}$ of the DC-DC converter, the DC-DC efficiency $\eta_{\text{4054}}$ of the STC4054 battery charging IC, and the DC-DC efficiency $\eta$ of the entire system, defined as the ratio of the power at the terminals of the Li-ion cell to the power at the USB connector of the charging station for different distances $h$ between the transmitting and receiving coils.}
    \label{fig:Efficiency}
\end{figure*}
%______________________Figure_S13_______________________

Although optimization of power transfer efficiency was not a primary goal of this work, detailed efficiency calculations throughout the power conversion chain have been performed for future reference. The total power transfer efficiency $\eta$ during charging is determined by losses in the charging station, WPT losses, losses in the rectifier, losses in the DC-DC converter, and, finally, dissipation in the STC4054 charging controller. Therefore, we estimate the total system efficiency as
\begin{equation}
    \eta = \eta_{\text{DC-AC}}\cdot\eta_{\text{AC-AC}}\cdot\eta_{\text{AC-DC}}\cdot\eta_{\text{DC-DC}}\cdot\eta_{4054},
\end{equation}
where $\eta_{\text{DC-AC}}$ is the efficiency of DC to AC conversion by the charging station, $\eta_{\text{AC-AC}}$ is the WPT efficiency, $\eta_{\text{AC-DC}}$ is the rectifier efficiency, $\eta_{\text{DC-DC}}$ is the DC-DC converter efficiency, and $\eta_{4054}$ is the STC4054 charging controller efficiency. Due to the compact size of the device, it is difficult to measure voltages, and especially currents, at each stage of the power chain directly in the experiment. Therefore, we perform experimental efficiency measurements of isolated stages wherever possible and estimate the efficiencies of the stages that are difficult to measure directly, taking into account the changes in their load impedances with the state of charge of the Li-ion cell.

We start by estimating the efficiency $\eta_{\text{DC-AC}}$ of the charging station, considering the DFRL~R2~v~5.0 module as an example. This module uses a full-bridge inverter based on an AO4803 dual p-MOSFET and an AO4882 dual n-MOSFET driving a serial $LC$ circuit consisting of a 400~nF 100~V polymer film capacitor, with an ESR of 20~mOhms at 100-200~kHz and a standard A11-type 42~mm coil with an inductance of 6~$\mu$H and ESR of 0.25~Ohm. The inverter current also passes through a 20~mOhm current sense resistor. The main power loss channels include dissipation at open MOSFETs, losses in the $LC$ circuit, and dissipation at the current sense resistor. By neglecting the power consumed by the driver chip, let us estimate the instantaneous voltage $V_{\text{coil}}$ and current $I_{\text{coil}}$ passing through the $LC$ circuit, assuming a 5~V supply voltage. The current $I_{\text{coil}}$ is preserved throughout this power chain, while
\begin{equation}
    V_{\text{coil}} = 5~\text{V} - I_{\text{coil}} \cdot \left(2R_{\text{ds}}^{\text{(on)}} + R_{\text{sense}} + \text{ESR}_{C} + \text{ESR}_{L} \right),
\end{equation}
where $2R_{\text{ds}}^{\text{(on)}} = 50~\text{mOhm}+20~\text{mOhm}$ is the sum of the on-state resistances of the p- and n-MOSFETs, $R_{\text{sense}} = 20~\text{mOhm}$ is the value of the current sense resistor, $\text{ESR}_{C} = 20~\text{mOhm}$ is the capacitor ESR, and $\text{ESR}_{L} = 250~\text{mOhm}$ is the coil ESR. Therefore, the efficiency $\eta_{\text{DC-AC}}$ of the charging station evaluates as
\begin{equation}
    \eta_{\text{DC-AC}} = 1 - \frac{I_{\text{coil}} \cdot 360~\text{mOhm}}{5~\text{V}}.
\end{equation}
Assuming that the charging station emits 5~W power, $I_{\text{coil}} \simeq 1~\text{A}$, and the efficiency of the station is at least as high as $\eta_{\text{DC-AC}} \simeq 93\%$.

To evaluate the efficiencies of each of the remaining stages of the power chain, we need to take into account their varying load impedances. They are ultimately determined by the differential resistance of the Li-ion cell
\begin{equation}
    Z_{\text{cell}} = \frac{\partial V_{\text{cell}}}{\partial I_{\text{cell}}},
\end{equation}
which changes depending on its state of charge (SoC). Figure~\ref{fig:Efficiency}(a) shows the time dependencies of the voltage $V_{\text{cell}}$ and the current $I_{\text{cell}}$ of the Li-ion cell as functions of time as it is charged with the STC4054 chip, as well as the corresponding SoC. We use these data to calculate the effective impedance $Z_{\text{cell}}$ of the Li-ion cell, which is shown in Figure~\ref{fig:Efficiency}(b) as a function of the SoC.

Next, each of the power conversion stages changes this impedance to 
\begin{equation*}
    Z_{\text{in}} = \frac{\partial V_{\text{in}}}{\partial I_{\text{in}}},
\end{equation*}
where $V_{\text{in}}$ is the input voltage of the corresponding power conversion stage and $I_{\text{in}}$ is its input current. For the STC4054 charge control chip, the input current is approximately equal to the battery current $I_{\text{cell}}$ (the currents drawn by the chip itself and the LED do not exceed 1~mA and are negligible), while its input voltage equals 5~V. Therefore,
\begin{equation*}
    Z_{4054}(\text{SoC}) = \frac{5~V}{I_{\text{cell}}(\text{SoC})}.
\end{equation*}
For the DC-DC converter, the input current $I_{\text{in}}^{\text{DC-DC}}(Z_{4054})$ of the DC-DC converter has been measured experimentally as a function of the DC-DC converter load $Z_{4054}$, while the input voltage is set to 7~V (the setpoint voltage maintained by the Qi protocol feedback loop). Therefore,
\begin{equation*}
    Z_{\text{DC-DC}}(\text{SoC}) = \frac{7~V}{I_{\text{in}}^{\text{DC-DC}}(Z_{4054}(\text{SoC}))}.
\end{equation*}
The efficiency of the DC-DC converter has been measured at the same time, and is shown in Figure~\ref{fig:Efficiency}(c).

Finally, the input impedance of the rectifier is approximately equal $Z_{\text{R}}\simeq 1.05 Z_{\text{DC-DC}}$, as the current is conserved throughout the rectifier, but there is a forward voltage drop of approximately 5\% of the input voltage amplitude (see the Supplementary Note~8). From the same considerations, we can readily approximate $\eta_{\text{AC-DC}}\simeq 95\%$. The input impedances obtained this way are shown in Fig.~\ref{fig:Efficiency}(b) and used in the efficiency calculations at each step of the power conversion chain. 

The efficiency $\eta_{\text{AC-AC}}$ of wireless power transmission is defined as the ratio of the AC powers on inductively coupled receiver and transmitter $LC$ circuits. It can be evaluated from the measurements of the $S$-parameters of the coupled serial contours as a function of the distance $h$ between the transmitting and receiving coils, as well as the function of SoC:
\begin{equation}
\label{eq:ACACeff}
    \eta_{\text{AC-AC}} = \frac{(1 - |\Gamma_{\text{L}}|^2) |S_{21}|^2}{
    | 1 - S_{22}\Gamma_{\text{L}} |^2
    - \left|
    S_{11} + \frac{S_{12}S_{21}\Gamma_{\text{L}}}{1 - S_{22}\Gamma_{\text{L}}}
    \right|
    \cdot | 1 - S_{22}\Gamma_{\text{L}} |^2,
}
\end{equation}
where
\begin{equation*}
    \Gamma_{\text{L}} = \frac{0.81\cdot Z_{\text{R}}(\text{SoC}) - 50}{0.81\cdot Z_{\text{R}}(\text{SoC}) + 50}.
\end{equation*}
The factor 0.81 in the AC load impedance as seen by the receiver $LC$ circuit is responsible for the RMS-to-amplitude conversion. We also ignore the imaginary part of the AC load impedance (related to the diode capacitance, inductance of the DC-DC converter, etc.), as it is small compared to the real part and difficult to estimate correctly.

The WPT efficiencies evaluated from the experimentally measured $S$-parameters using Eq.~\eqref{eq:ACACeff} are shown in Fig.~\ref{fig:Efficiency}(d) for $f = 200~\text{kHz}$ and decrease with both the state of charge and the distance $h$. The maximum achievable efficiency observed for $h=0$ reaches 51\%, which decreases to 19\% for $\text{SoC} = 100\%$. For $h = 10~\text{mm}$, the efficiency starts at 10\% at the beginning of the charging process and drops to 3\% at the end of the charging process. The decrease in efficiency with an increasing state of charge is explained by an increasing impedance mismatch.

The second plot in Fig.~\ref{fig:Efficiency}(d) shows the efficiency of the DC-DC converter $\eta_{\text{DC-DC}}$ as a function of the state of charge, which is observed to start at 80\% at the beginning of the charge process and decreases to 40\% at its end due to the impedance of the Li-ion cell being non-optimal for the DC-DC converter. The third plot in Fig.~\ref{fig:Efficiency}(d) shows the efficiency $\eta_{4054}$ of the STC4054 charging controller as a function of the state of charge. As STC4054 is a dissipative controller, its efficiency is the lowest at the beginning of the charging process (65\%), when the voltage drop across the chip is the greatest, and gradually increases to 83\% as the Li-ion cell voltage increases so the voltage drop across the chip becomes lower.

Finally, the fourth plot in Fig.~\ref{fig:Efficiency}(d) shows the total system efficiency $\eta$ measured experimentally as a function of SoC for distances $h$ from 0 to 7~mm. To eliminate SoC drift during the experiment, the Li-ion cell was replaced with resistors with values determined as $Z_{\text{cell}}(\text{SoC})$ in this experiment. The DC voltage and current were then measured at the input of the charging station to obtain the input power, and the DC voltage across the battery equivalent $Z_{\text{cell}}(\text{SoC})$ was measured to obtain the output power. It is seen that the dependencies $\eta(h, \text{SoC})$ follow the same tendencies as $\eta_{\text{AC-AC}}(h, \text{SoC})$, but are lower due to the losses in the charger, rectifier, DC-DC converter, and the STC4054 charging controller. These losses can potentially be alleviated by eliminating the DC-DC converter and the charging controller in favor of using the feedback loop of the Qi protocol to maintain the necessary $V_{\text{cell}}$ directly at the rectifier output. However, guaranteed prevention of cell overvoltage without dissipating excess energy as heat becomes an intractable issue in this case, as Qi charging stations may, under certain conditions, generate voltages up to $12~\text{VDC}$ at the output of the rectifier before a power transfer contract is established and the feedback loop is closed.

%__________________________S12_Comparison_____________________________
\section*{Supplementary Note 12. Comparison with conventional AA batteries}

The comparison of the characteristics of the prototype demonstrated in the current work with conventional AA batteries is summarized in Table~\ref{tab:AAbatteries_conv}. It is seen that the introduced wirelessly charged AA battery prototype has a number of charging cycles comparable to other rechargeable batteries, as follows from the characteristics of the Li-ion cell, but a considerably lower capacity. On the other hand, the weight of the prototype is also lower compared to that of other batteries. The reason for both differences is that we use a compact Li-ion cell that can be replaced by a one with higher capacity (and weight) once the PCB of the prototype is replaced by a more compact version based on the application-specific integrated circuit.

%_________________________Table_S1___________________________
\begin{table}[hbtp]
    \centering
    \caption{Comparison with traditional AA batteries by basic characteristics.}
    \begin{tabular}{| l | c | c | c | c | c | c |}
    \hline
    & Zinc-carbon & Alkaline & Ni-Cd & NiMH & Li-ion 1.5V & Wireless (this work) \\\hline
    Weight, g & 14-19 & 23 & 17-23 & 25-30 & 16-19 & 9 \\
    Number of charging cycles & 1 & 1 & 500-1000 & 500-1000 & 500-1000 & 800 \\
    Capacity, Wh & 0.6-1.5 & 2-4 & 0.6-1.2 & 2-3 & 2-3.5 & 0.24 (prototype) \\
    \hline
    \end{tabular}
    \label{tab:AAbatteries_conv}
\end{table}
%_________________________Table_S1___________________________

\end{spacing}